\documentclass[twocolumn]{aastex63}

\usepackage{amsmath}

 \newcommand{\E}[1]{\ensuremath{\times 10^{#1}}}

\def\gtrsim{\mathrel{\hbox{\rlap{\hbox{\lower4pt\hbox{$\sim$}}}\hbox{$>$}}}}

\def \arcsec {\hbox{$^{\prime\prime}$}}

\received{\today}
\revised{--}
\accepted{--}

\submitjournal{ApJ}

\shorttitle{J18219}
\shortauthors{O'Connor et al.}
\graphicspath{{./}{figures/}}

\begin{document}

\title{Identification of an X-ray Pulsar in the BeXRB system IGR J18219$-$1347}

\correspondingauthor{Brendan O'Connor}
\email{oconnorb@gwmail.gwu.edu}

\author[0000-0002-9700-0036]{B. O'Connor}
    \affiliation{Department of Physics, The George Washington University, Washington, DC 20052, USA}
    \affiliation{Astronomy, Physics and Statistics Institute of Sciences (APSIS), The George Washington University, Washington, DC 20052, USA}
    \affiliation{Department of Astronomy, University of Maryland, College Park, MD 20742-4111, USA}
    \affiliation{Astrophysics Science Division, NASA Goddard Space Flight Center, 8800 Greenbelt Rd, Greenbelt, MD 20771, USA}
\author[0000-0002-5274-6790]{E. G\"{o}\u{g}\"{u}\c{s}}
    \affiliation{Sabanc\i~University, Faculty of Engineering and Natural Sciences, \.Istanbul 34956 Turkey}
\author[0000-0002-1169-7486]{D. Huppenkothen}
    \affiliation{SRON Netherlands Institute for Space Research, Niels Bohrweg 4, 2333CA Leiden, The Netherlands}
\author[0000-0003-1443-593X]{C. Kouveliotou}
    \affiliation{Department of Physics, The George Washington University, Washington, DC 20052, USA}
    \affiliation{Astronomy, Physics and Statistics Institute of Sciences (APSIS), The George Washington University, Washington, DC 20052, USA}
\author[0000-0002-1653-6411]{N. Gorgone}
    \affiliation{Department of Physics, The George Washington University, Washington, DC 20052, USA}
    \affiliation{Astronomy, Physics and Statistics Institute of Sciences (APSIS), The George Washington University, Washington, DC 20052, USA}
\author[0000-0001-9788-3345]{L.~J. Townsend}
    \affiliation{South African Astronomical Observatory, P.O. Box 9, 7935 Observatory, South Africa}
\author[0000-0002-0882-7702]{A. Calamida}
    \affiliation{Space Telescope Science Institute, 3700 San Martin Drive, Baltimore, MD 21218, USA}
\author[0000-0002-6652-9279]{A. Fruchter}
    \affiliation{Space Telescope Science Institute, 3700 San Martin Drive, Baltimore, MD 21218, USA}
\author[0000-0002-7004-9956]{D.~A.~H. Buckley}
    \affiliation{Department of Astronomy, University of Cape Town, Private Bag X3, Rondebosch 7701, South Africa}
    \affiliation{South African Astronomical Observatory, P.O. Box 9, 7935 Observatory, South Africa}
\author[0000-0003-4433-1365]{M.~G. Baring}
    \affiliation{Department of Physics and Astronomy - MS 108, Rice University, 6100 Main Street, Houston, Texas 77251-1892, USA}
\author[0000-0002-6745-4790]{J.~A. Kennea}
    \affiliation{Department of Astronomy and Astrophysics, The Pennsylvania State University, 525 Davey Lab, University Park, PA 16802, USA}
\author[0000-0002-7991-028X]{G. Younes}
    \affiliation{Department of Physics, The George Washington University, Washington, DC 20052, USA}
    \affiliation{Astronomy, Physics and Statistics Institute of Sciences (APSIS), The George Washington University, Washington, DC 20052, USA}

\author{Z. Arzoumanian}
    \affiliation{X-Ray Astrophysics Laboratory, NASA Goddard Space Flight Center, 8800 Greenbelt Rd, Greenbelt, MD 20771, USA}
\author[0000-0001-8018-5348]{E. Bellm} 
    \affiliation{DIRAC Institute, Department of Astronomy, University of Washington, 3910 15th Avenue NE, Seattle, WA 98195, USA}
\author[0000-0003-1673-970X]{S.~B. Cenko}
    \affiliation{Astrophysics Science Division, NASA Goddard Space Flight Center, 8800 Greenbelt Rd, Greenbelt, MD 20771, USA}
    \affiliation{Joint Space-Science Institute, University of Maryland, College Park, MD 20742 USA}
\author[0000-0001-7115-2819]{K. Gendreau}
    \affiliation{X-Ray Astrophysics Laboratory, NASA Goddard Space Flight Center, 8800 Greenbelt Rd, Greenbelt, MD 20771, USA}
\author[0000-0001-8530-8941]{J. Granot}
    \affiliation{Department of Physics, The George Washington University, Washington, DC 20052, USA}
    \affiliation{Department of Natural Sciences, The Open University of Israel, P.O Box 808, Ra'anana 43537, Israel}
    \affiliation{Astrophysics Research Center of the Open university (ARCO), The Open University of Israel, P.O Box 808, Ra’anana 43537, Israel}
\author{C. Hailey}
    \affiliation{Columbia Astrophysics Laboratory, Columbia University, New York, NY 10027, USA}
\author[0000-0003-2992-8024]{F. Harrison}
    \affiliation{Cahill Center for Astrophysics, California Institute of Technology, 1216 East California Boulevard, Pasadena, CA 91125, USA}
\author[0000-0002-8028-0991]{D. Hartmann}
    \affiliation{Department of Physics and Astronomy, Clemson University, Kinard Lab of Physics, Clemson, SC 29634-0978, USA}
\author{L. Kaper}
    \affiliation{University of Amsterdam, Science Park 904, 1098 XH Amsterdam, The Netherlands}
\author[0000-0002-2715-8460]{A. Kutyrev}   
    \affiliation{Department of Astronomy, University of Maryland, College Park, MD 20742-4111, USA}
    \affiliation{Astrophysics Science Division, NASA Goddard Space Flight Center, 8800 Greenbelt Rd, Greenbelt, MD 20771, USA}
\author[0000-0002-6986-6756]{P.~O. Slane}
    \affiliation{Center for Astrophysics, Harvard \& Smithsonian, 60 Garden St. Cambridge, MA 02138, USA}
\author[0000-0003-2686-9241]{D. Stern}
    \affiliation{Jet Propulsion Laboratory, California Institute of Technology, 4800 Oak Grove Drive, Mail Stop 169-221, Pasadena, CA 91109, USA}
\author[0000-0002-1869-7817]{E. Troja}
    \affiliation{Department of Astronomy, University of Maryland, College Park, MD 20742-4111, USA}
    \affiliation{Astrophysics Science Division, NASA Goddard Space Flight Center, 8800 Greenbelt Rd, Greenbelt, MD 20771, USA}
\author[0000-0001-9149-6707]{A.~J. van der Horst}
    \affiliation{Department of Physics, The George Washington University, Washington, DC 20052, USA}
   \affiliation{Astronomy, Physics and Statistics Institute of Sciences (APSIS), The George Washington University, Washington, DC 20052, USA}
\author[0000-0002-3101-1808]{R.~A.~M.~J. Wijers}
    \affiliation{Anton Pannekoek Institute, University of Amsterdam, Postbus 94249, 1090 GE Amsterdam, The Netherlands}
\author[0000-0002-6896-1655]{P. Woudt}
    \affiliation{Department of Astronomy, University of Cape Town, Private Bag X3, Rondebosch 7701, South Africa}



\begin{abstract}
We report on observations of the candidate Be/X-ray binary IGR J18219$-$1347 with \textit{Swift}/XRT, \textit{NuSTAR}, and \textit{NICER} during Type-I outbursts in March and June 2020. Our timing analysis revealed the spin period of a neutron star with $P_\textrm{spin}=52.46$ s. This periodicity, combined with the known orbital period of $72.4$ d, indicates that the system is a BeXRB. Furthermore, by comparing the infrared counterpart's spectral energy distribution to known BeXRBs, we confirm this classification and set a distance of approximately $10-15$ kpc for the source.  The source's broadband X-ray spectrum ($1.5-50$ keV) is described by an absorbed power-law with photon index $\Gamma$\,$\sim$\,$0.5$ and cutoff energy at $\sim$\,$13$ keV.
\end{abstract}

\keywords{
X-ray astronomy (1810) --- High Mass X-ray Binary stars (733)  --- Pulsars (1306) --- Neutron Stars (1108)}


\section{Introduction}
\label{sec: intro}








High-mass X-ray binaries (HMXB) comprise a compact object (white dwarf, neutron star (NS) or black hole) and a massive ($> 10 M_\odot$) companion star, donating matter to it. A sub-class of HMXBs, known as Be/X-ray binaries (BeXRB), consist of a compact object with a Be star companion with a decretion disk, which is formed by material ejected from the Be star's surface due to its rapid rotation (see \citealt{Rivinius2013} for a recent review).
BeXRBs make up to $\sim$\,$49\%$ of the HMXB population in the Milky Way \citep{Coleiro2013}. 

Accretion occurs as the compact object, primarily a NS, which is generally on a wide, highly eccentric orbit, passes through the decretion disk of the Be companion. During these passages, the system undergoes periodic bright \textit{Type I} outbursts  \citep[lasting days to weeks;][]{Okazaki2001,Reig2007,Chaty2011}. BeXRBs  generally exhibit long orbital periods \citep[$15$ to $400$ d;][]{Reig2011}, which are found to be correlated to the spin period of the compact object \citep[cf. Corbet Diagram;][]{Corbet1984,Corbet1986}. 
Long-term monitoring is critical for uncovering the binary orbital period, confirmed through the repeated detection of \textit{Type I} outbursts.

IGR J18219$-$1347 (hereafter J18219) was discovered with the \textit{INTEGRAL} satellite in 2010 \citep{Krivonos2010}. An earlier X-ray analysis of \textit{Swift}/BAT and XRT data \citep{LaParola2013} showed that the source X-ray flux exhibited strong  variability as a function of its orbit, leading to periodic outbursts. \citet{LaParola2013} associated these with the periastron passage of the compact object, leading to the determination of an orbital period of $\sim$\, $72.4$ d. Further evidence of the BeXRB nature of the system was reported by \citet{Karasev2012}. Their \textit{Chandra} localization of the source coincided with a bright infrared (IR) counterpart in the UKIRT Infrared Deep Sky Survey \citep[UKIDSS;][]{Lawrence2007}; a candidate Be star.


We detected J18219 during our \textit{Swift} Deep Galactic Plane Survey (DGPS; PI: C. Kouveliotou).  We present here new X-ray observations of the source obtained with \textit{Swift}, \textit{NICER}, and \textit{NuSTAR}. We organize the paper as follows. We introduce the observations and data analysis in \S \ref{sec: obs/analysis}. In \S \ref{sec: results}, we report on the timing and spectral analyses of our X-ray data and on our search for the optical counterpart of J18219. Finally, we compare the candidate IR counterpart spectral energy distribution (SED) to known Be stars (\S \ref{sec: counterpart}). We present a discussion of our results in \S \ref{sec: discussion} and our conclusions in \S \ref{sec: conclusions}.

Unless otherwise stated, confidence intervals/upper limits are presented at the $1\sigma$/$3\sigma$ level, respectively. Photometry is reported in the AB magnitude system, except where specified differently. 

\section{Observations and Data Analysis}
\label{sec: obs/analysis}

We detected J18219 in March 2020 with the \textit{Neil Gehrels Swift} Observatory \citep{Gehrels2004} X-ray Telescope \citep[XRT;][]{Burrows2005} in Photon Counting (PC) mode. The source brightness justified triggering our approved Target of Opportunity (ToO) observation with the Nuclear Spectroscopic Telescope ARray \citep[\textit{NuSTAR};][]{Harrison+2013}. We observed the source again in May 2020 to complete the required DGPS 5\,ks exposure of the tile. The source brightness indicated a possible outburst, leading to a Neutron Star Interior Composition Explorer \citep[\textit{NICER};][]{Gendreau2016} Director's Discretionary Time (DDT) request. Table \ref{tab:Xobservations} shows the log of all X-ray observations.

We also performed optical imaging with the Robert Stobie Spectrograph (RSS) on the 11-m Southern African Large Telescope (SALT) and the Large Monolithic Imager (LMI) on the 4.3-m Lowell Discovery Telescope (LDT) to identify and characterize the optical counterpart of J18219 (see \S \ref{sec: opt_imaging}). In addition, we analyzed archival UKIDSS infrared imaging.


\subsection{X-ray Observations}

\subsubsection{\textit{Swift}/BAT}
\label{sec: BAT}



J18219 is one of the long-term monitoring targets with the \textit{Swift}/Burst Alert Telescope \citep[BAT;][]{Barthelmy2005}. All target data are daily averaged in the $15-50$\,keV energy band and stored at the \textit{Swift}/BAT Hard X-ray Transient Monitor archive\footnote{\url{https://swift.gsfc.nasa.gov/results/transients/weak/SWIFTJ1821.8-1348/}} \citep{Krimm2013}. We analyzed data spanning 3369 days (MJD $55968-59337$) to refine the orbital period previously identified by \citet{LaParola2013} (see \S \ref{sec: timing}). 
The data were not barycenter corrected; the significantly long orbital period (72.4 d) renders the correction effect negligible. 

\subsubsection{\textit{Swift}/XRT}
\label{sec: XRT}


\textit{Swift}/XRT observations of J18219 comprise 22 epochs, totaling 49.4 ks, with 29.8 ks in Windowed Timing (WT) mode and 19.6 ks in PC mode. The WT mode data comprise largely the observing campaign requested by \citealt{Krimm2012} and reported by \citet{LaParola2013} (see Table \ref{tab:Xobservations}). In this work we analyze all WT data together with the PC mode observations (ObsIDs: 3110746, 3110747, 3110855) obtained through the DGPS.

We reduced and analyzed the PC mode observations using standard filtering and cleaning procedures in the \texttt{xrtpipeline} software. The source count rates were determined using the \texttt{ximage} routine \texttt{sosta} within \texttt{HEASoft v6.27.2}. We utilized source extraction regions corresponding to an 87\% enclosed-energy fraction, and local background annuli surrounding these regions. We then corrected the count rates for vignetting, bad pixels/columns on the CCD, and point-spread function (PSF) losses, using the \texttt{xrtmkarf} command combined with the exposure map to recover the full 100\% of the enclosed-energy fraction. 

Finally, we used the {\it Swift}/XRT data products generator\footnote{\url{https://www.swift.ac.uk/user_objects/}} to obtain the most accurate source position based on all PC mode exposures. The XRT enhanced position \citep{Evans2009} is RA, DEC (J2000) = $18^{h}21^m 54^{s}.92$, $-13^\circ 47\arcmin 23.3\arcsec$ with an accuracy of $3.5\arcsec$ (90\% confidence level; hereafter CL). This is consistent with the \textit{Chandra} localization reported by \citet{Karasev2012}: RA, DEC (J2000) = $18^{h}21^m 54^{s}.821$, $-13^\circ 47\arcmin 26.703\arcsec$ with uncertainty $0.9\arcsec$ (90\% CL). 

\subsubsection{\textit{NuSTAR}}
\label{sec: nustar}


We used one of our {\it NuSTAR} ToOs to observe J18219 on March 15, 2020 for 23 ks (ObsID: 90601309002).
\textit{NuSTAR} comprises two identical focal plane modules, FPMA and FPMB, covering $3-79$ keV.
The data reduction was performed using the \textit{NuSTAR} Data Analysis Software pipeline (\texttt{NuSTARDAS}) v1.9.2 and the calibration files (CALDB) version 20200726 within \texttt{HEASoft v6.27.2}. The data were processed using \texttt{nupipeline}, and then lightcurves and spectra were extracted using \texttt{nuproducts}.
Source spectra were extracted from a 100\arcsec{} radius region centered on the transient. The background was similarly extracted from a 100\arcsec{} radius source-free region. For our spectral analysis we truncated the \textit{NuSTAR} data at 50~keV, where the background began to dominate; the spectra were grouped to a minimum of 25 counts per bin for statistical significance. The photon arrival times were barycenter-corrected to the solar system using the \texttt{barycorr}\footnote{\url{https://heasarc.gsfc.nasa.gov/ftools/caldb/help/barycorr.html}} tool and the latest clockfile\footnote{nuCclock20100101v116.fits.gz, see \url{https://nustarsoc.caltech.edu/NuSTAR_Public/NuSTAROperationSite/clockfile.php}}.

We note that in FPMB the source position is partially contaminated by stray light from the bright low-mass X-ray binary (LMXB) GX 17+2 ( \citealt{straycats}). We chose, therefore, to perform the majority of our analysis using the uncontaminated FPMA data. For the FPMB data we carefully selected the background region to subtract and minimize the effects of the stray light; in all these cases we confirmed that including FPMB data did not change our results.



\subsubsection{\textit{NICER}}
\label{sec: nicer}

We observed J18219 with {\it NICER} on June 3, 2020 for 2.3 ks (ObsID: 3201610101) through a DDT request. 
The data were processed using \texttt{NICERDAS} v7a within \texttt{HEASoft v6.27.2} and filtered using standard cleaning criteria with \texttt{nicerl2}. The cleaned event file was barycenter-corrected (using \texttt{barycorr}) to the Solar System based on the \textit{Chandra} position. We then used the \texttt{xselect} task to extract the lightcurve and spectrum between $1-10$ keV.  The \textit{NICER} background spectrum was estimated using the \texttt{nibackgen3C50 v6} tool \citep{Remillard2021}; it dominates at $\lesssim 1.5$ keV, therefore, we exclude these energies from our spectral analysis. Finally, the spectra were grouped to a minimum of 25 counts per bin using \texttt{grppha}.

We carried out additional \textit{NICER} DDT observations on May 2, 2021 for 1.2 ks (ObsID: 4201610101). The source was not detected, and we adopt a $3\sigma$ upper limit ($0.4-12$ keV) of  $\sim$\,$1.2$ cts s$^{-1}$ \citep{Remillard2021}, which corresponded to an unabsorbed flux $\lesssim 1.5\times 10^{-11}$ erg cm$^{2}$ s$^{-1}$ for the best fit model spectrum. 




\subsection{Optical Imaging}
\label{sec: opt_imaging}

\subsubsection{Southern African Large Telescope (SALT)}

We carried out optical imaging with the Robert Stobie Spectrograph \citep[RSS;][]{Burgh2003,Kobulnicky2003,Smith2006} mounted on the 11-m SALT \citep{Buckley2006} on June 6, 2021. The observations were performed with a clear, fused silica filter for a total exposure time of 720 s. The data were processed by an automated SALT pipeline. We corrected the astrometry using the \texttt{astrometry.net} software \citep{Lang2010}. The seeing during these observations was very poor and no optical counterpart was identified at the \textit{Chandra} localization in the stacked image. The $3\sigma$ upper limit of the image is $\sim$\,$22.5$ AB mag.


\subsubsection{Lowell Discovery Telescope (LDT)}
\label{sec: LDT}

We performed optical observations with the Large Monolithic Imager (LMI) mounted on the 4.3-meter LDT (formerly the Discovery Channel Telescope) in Happy Jack, AZ on August 6, 2021 in the $i$ and $z$ filters for a total exposure of 1650 and 1000 s, respectively. 
The observations were performed under clear observing conditions with seeing $\sim$\,$1.25\arcsec$. The median airmass of the observations was $\sim$\,$1.5$. 

The data were reduced and analyzed using a custom pipeline \citep{Toy2016} which makes use of standard CCD reduction techniques in the \texttt{IRAF}\footnote{\texttt{IRAF} is distributed by the National Optical Astronomy Observatory, which is operated by the Association of Universities for Research in Astronomy (AURA) under cooperative agreement with the National Science Foundation (NSF).} package. We used \texttt{SCAMP} \citep{Bertin2006}  to align the individual frames, and then \texttt{SWarp} \citep{Bertin2010} to combine the images. The absolute astrometry was calibrated against the Panoramic Survey Telescope and Rapid Response System \citep[\textit{Pan-STARRS}, hereafter PS1;][]{Chambers2016,Flewelling2016} catalog. At the \textit{Chandra} source position, we do not detect the optical counterpart in either filter. 
The photometry was computed using the \texttt{SExtractor} \citep{Bertin1996} package, and was calibrated against stars in the PS1 Catalog. We obtained upper limits $i\gtrsim 23.7$ and $z\gtrsim 22.1$ AB mag at the source position (not corrected for Galactic extinction; see Table \ref{tab: optcounterpart}). 


\subsubsection{Pan-STARRS}
\label{sec: PS1}


We searched archival observations\footnote{\url{https://ps1images.stsci.edu/cgi-bin/ps1cutouts}} from PS1 \citep {Chambers2016,Flewelling2016} for the optical counterpart to J18219. At the \textit{Chandra} position we do not identify an optical source in any filter. We derive $3\sigma$ upper limits in the $g$, $r$, $i$, $z$, and $y$-bands. This photometry is reported in Table \ref{tab: optcounterpart}.

\subsubsection{Zwicky Transient Facility (ZTF)}

We analyzed public archival observations obtained with ZTF \citep{Bellm2019,Graham2019} between March 2018 and June 2021. The data were retrieved from The ZTF Image Service\footnote{\url{https://irsa.ipac.caltech.edu/Missions/ztf.html}} \citep{Masci2019}. We used \texttt{SWarp} to coadd all the individual science frames, each with an exposure time of 30 s, covering the position of J18219 in $g$ and $r$-band. This resulted in a total exposure of 3150 s (105 frames) and 8160 s (272 frames) in $g$ and $r$, respectively. In the $i$-band, due to the lack of publicly available observations, we make use of the reference image provided by ZTF, which comprises 15 stacked frames for a total of 450 s exposure. At the position of the infrared counterpart we do not detect a source to a depth $g\gtrsim 22.0$, $r\gtrsim 22.5$, and $i\gtrsim 20.9$ AB mag ($3\sigma)$. These limits are reported in Table \ref{tab: optcounterpart}, and are consistent with those derived from the PS1 (see \S \ref{sec: PS1}) and LDT imaging (\S \ref{sec: LDT}). 

\subsubsection{UKIDSS}
\label{sec: ukidss}

\begin{figure*}
\begin{tabular}{ccc}
\hspace{-0.5cm}
\includegraphics[width = 2.2in]{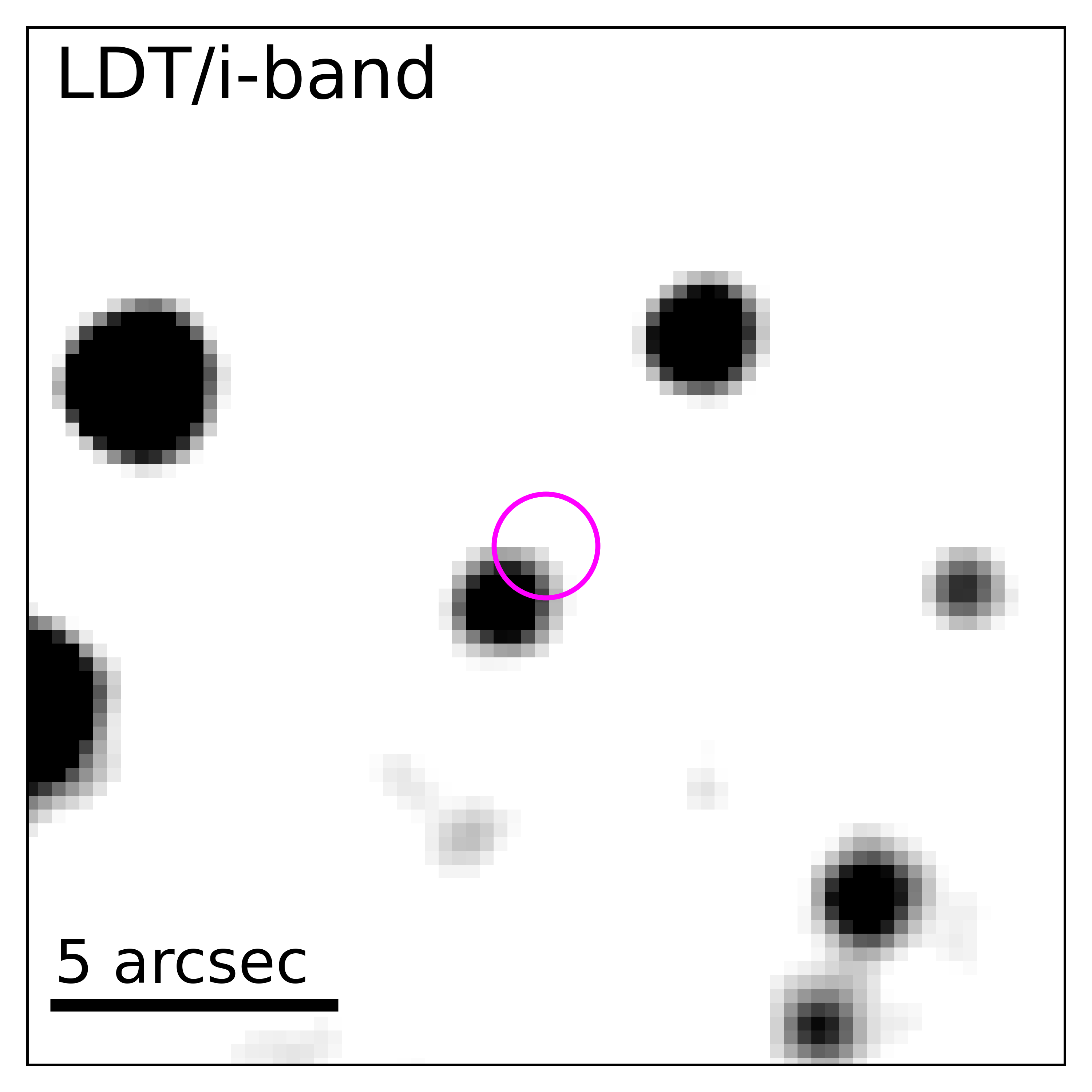} \hspace{-0.5cm}  &
\includegraphics[width = 2.2in]{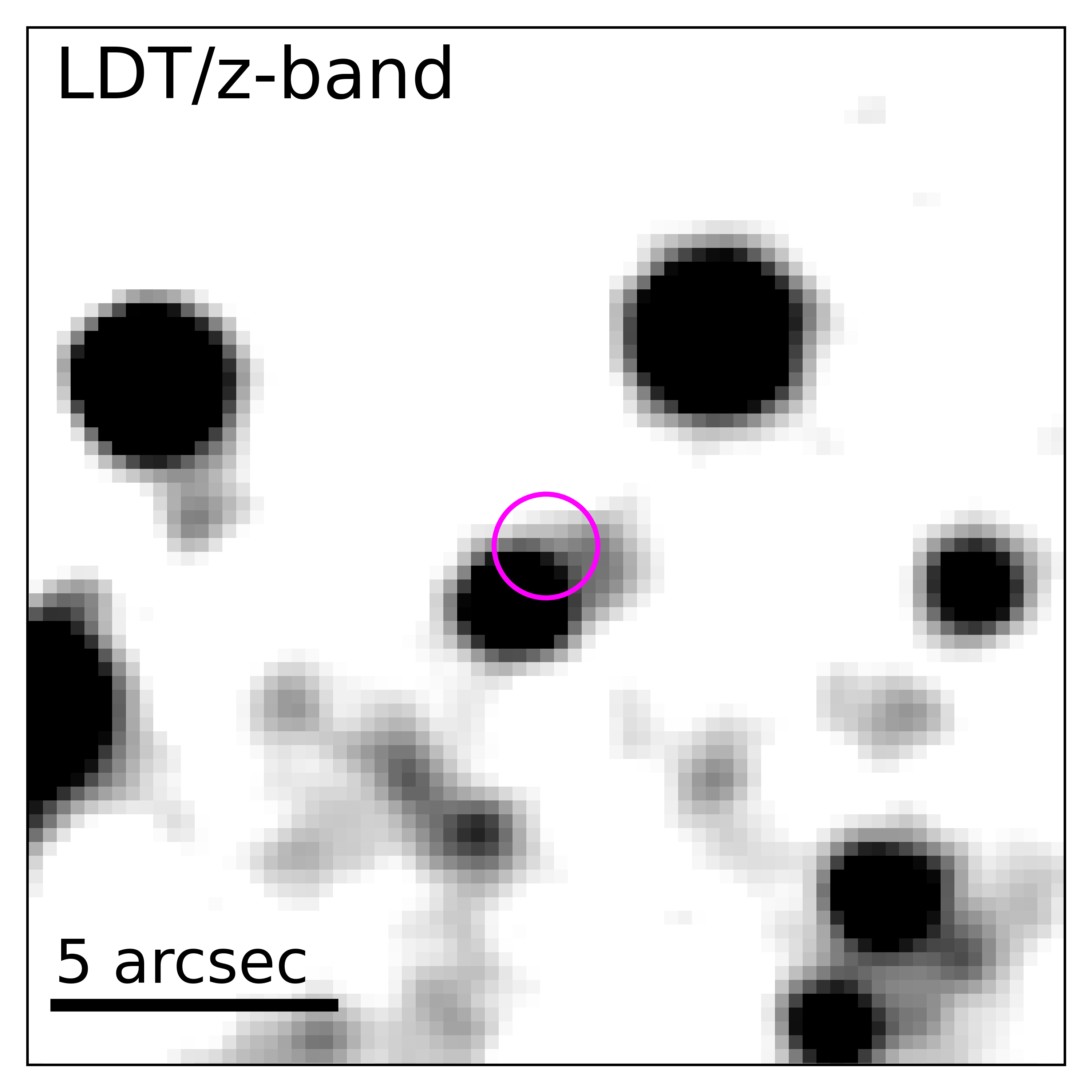} \hspace{-0.5cm} &
\includegraphics[width = 2.2in]{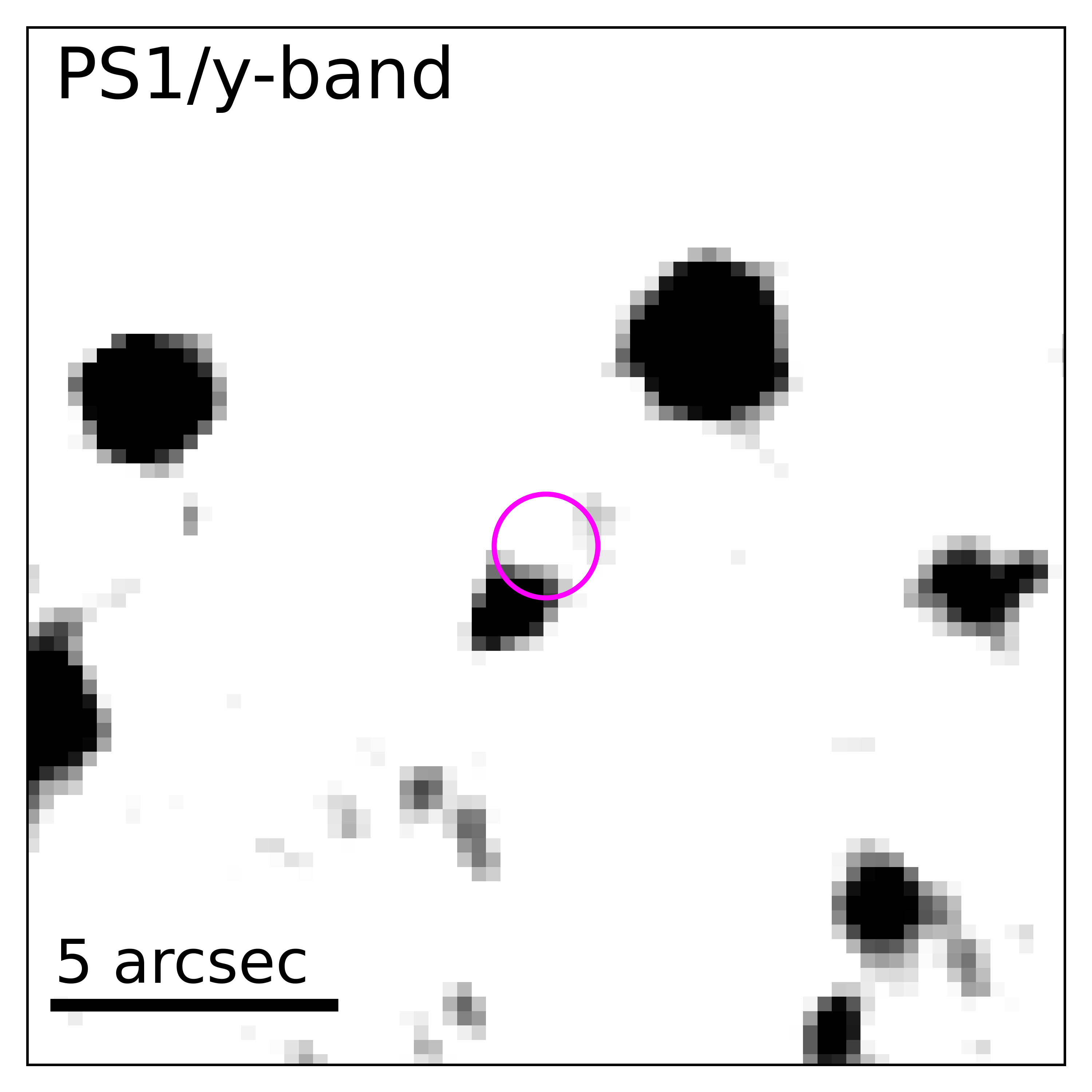} \\
\vspace{-0.5cm}
\hspace{-0.5cm}
\includegraphics[width = 2.2in]{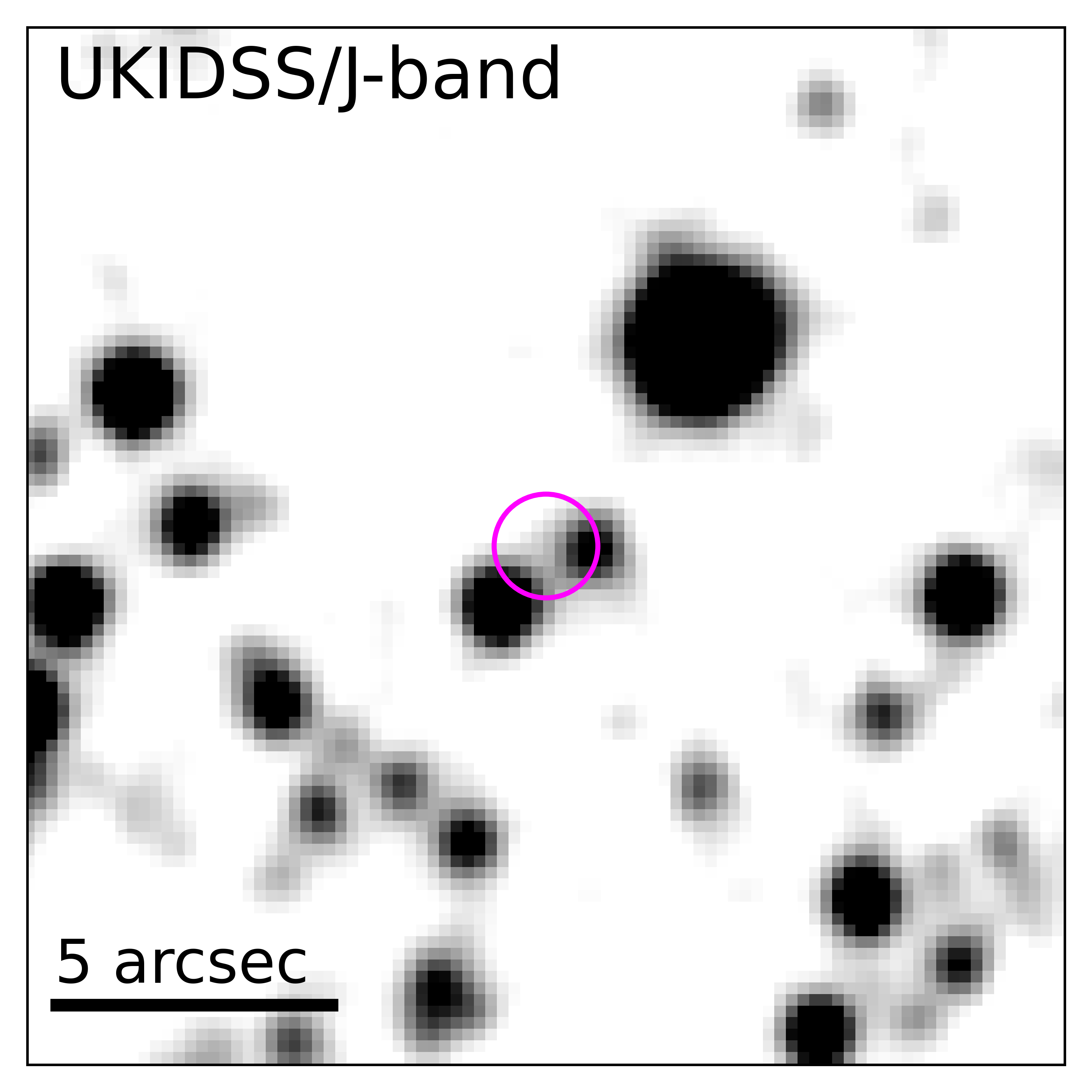} \hspace{-0.5cm}  &
\includegraphics[width = 2.2in]{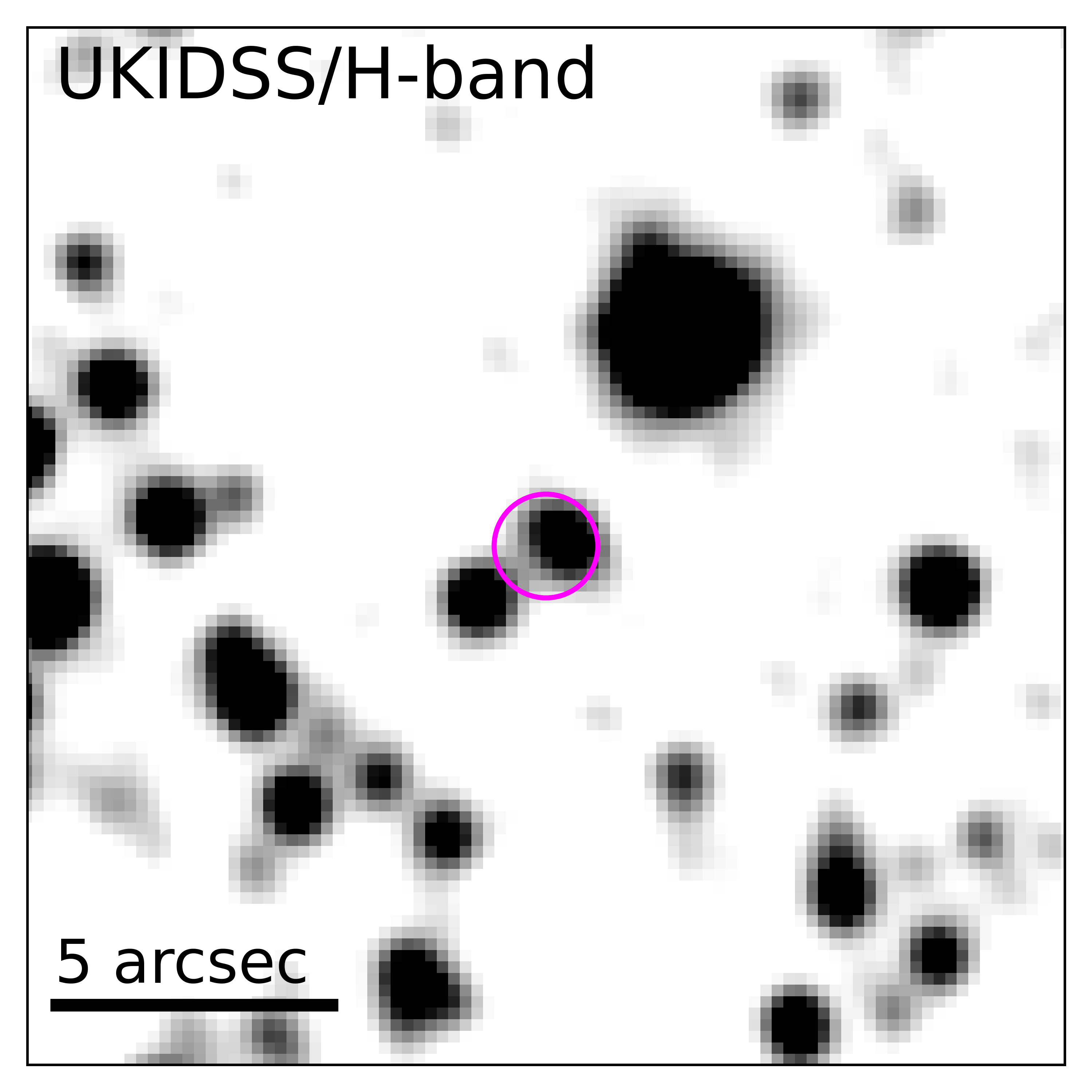} \hspace{-0.5cm} &
\includegraphics[width = 2.2in]{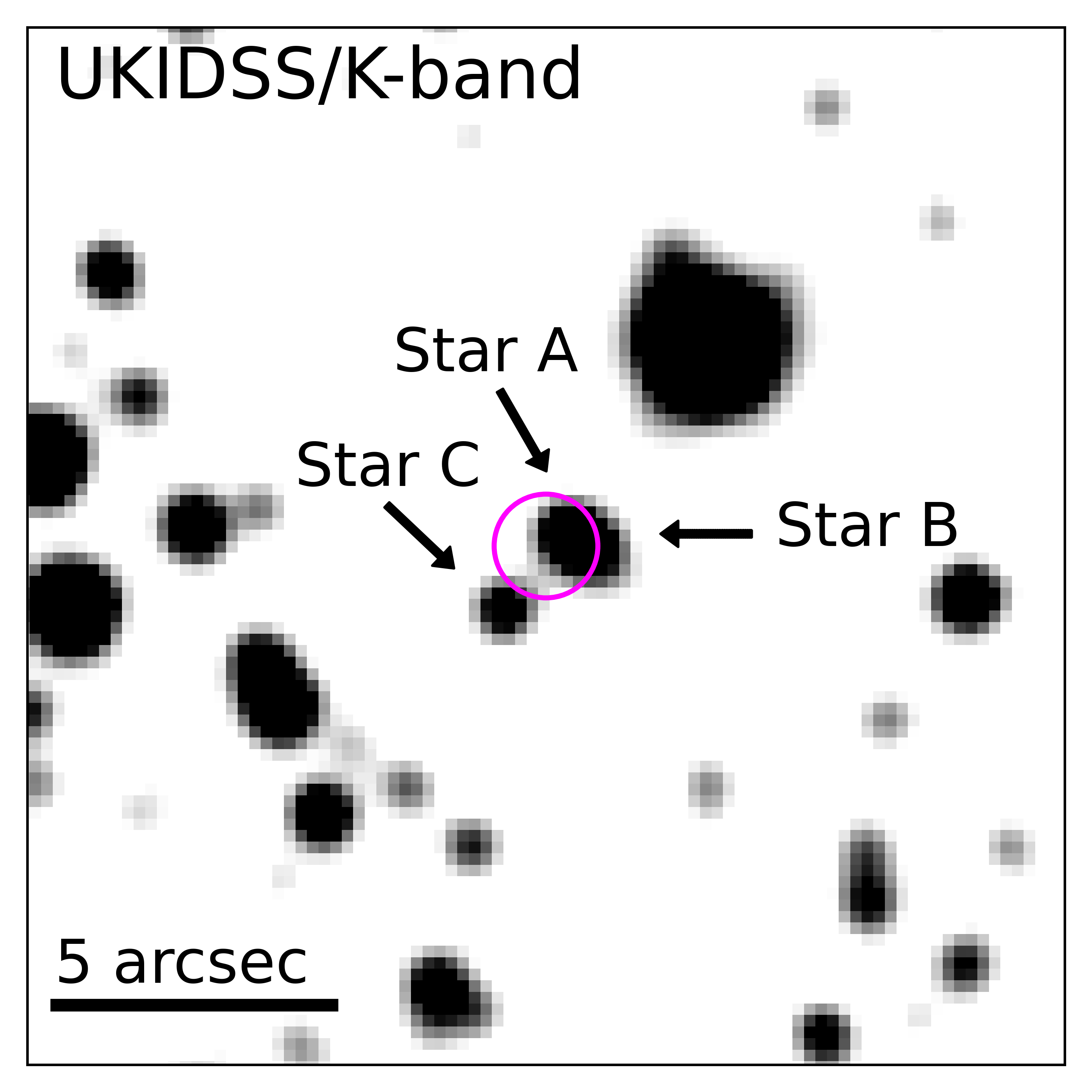} \\
\vspace{0.2cm}
\end{tabular}
\caption{The field of J18219 from our LDT imaging and archival observations from PS1 and UKIDSS. The \textit{Chandra}/HRC-I localization of the X-ray counterpart \citep{Karasev2012} is displayed by a magenta circle, of radius $0.9\arcsec$ (90\% CL). The position of the blended optical counterpart (Stars A and B) and the nearby Star C are labeled only in the UKIDSS/$K$-band figure (bottom right). Star A is the left star in the Star A/B complex.
In each figure, North is up and East is to the left. The images are smoothed for display purposes.
 \label{fig: opt_finding_chart} }
\end{figure*}

We analyzed public archival observations from UKIDSS \citep[DR11;][]{Lawrence2007} obtained in the $JHK$ filters with the Wide Field Camera \citep[WFCAM;][]{Casali2007} mounted on the 3.8-m United Kingdom Infrared Telescope (UKIRT). We downloaded the calibrated images from the WFCAM science archive \citep{Hambly2008}, which showed that the immediate field surrounding J18219 is relatively sparse (Figure \ref{fig: opt_finding_chart}). We identified in these images the infrared counterpart of J18219 proposed by \citet{Karasev2012}. Despite the good seeing ($\sim$\,$0.6-0.7\arcsec$), this source appeared to be the blended combination of two point sources, specifically in the $H$ and $K$ filters, thus prohibiting the true counterpart identification.

To de-blend the photometry and resolve the individual sources, referred to as Star A and Star B, we first performed PSF photometry with \texttt{DAOPHOT IV/ALLSTAR} \citep{Stetson1987}. We also identified another star, referred to as Star C, which lies just outside the \textit{Chandra} localization (90\% CL), and could, therefore, also be considered a potential counterpart. 

Next, we generated a photometric catalog for each of the three images (one per filter): the sources were identified in the $K$-band image. We used this image to create a list of objects in the field of view and performed forced photometry on all images with \texttt{ALLFRAME} \citep{Stetson1994} by using the previously calculated PSFs. These are Moffat functions with a quadratic spatial variation in the field. In order to improve the sky background estimate and the signal-to-noise ($S/N$) ratio we re-calculated the PSF for each image by using the output \texttt{ALLFRAME} catalogs and re-run the forced photometry. To calibrate the photometry to the Vega system we normalized for exposure time, calculated an aperture correction and used the zero points provided by CASU\footnote{\url{http://casu.ast.cam.ac.uk/surveys-projects/wfcam/technical/photometry}}.
The final catalog includes a total of 4,172 stars, out of which 3,067 have a measurement in $JHK$; it reaches $S/N \approx$ 5 at $K \approx$ 18.5 Vega mag (20.4 AB mag). 

Finally, we converted the photometry from the Vega to the AB magnitude system by using the definition in \citet{Hewett2006}. The final calibrated photometry for both Star A and Star B is tabulated in Table \ref{tab: optcounterpart}. We discuss these results in \S \ref{sec: counterpart}.


\begin{table}
    \centering
    \caption{Photometry of the optical/infrared counterparts (Stars A and B) of J18219. The photometry $m_\lambda$ is not corrected for Galactic extinction $A_\lambda$ due to interstellar reddening $E(B-V)=9.16$ mag \citep{Schlafly2011} in the direction of the source. The magnitudes $m_\lambda$ are reported in the AB magnitude system. 
    }
    \label{tab: optcounterpart}
    \begin{tabular}{lcccc}
    \hline
    \hline
\textbf{Source} & \textbf{Filter}  & \multicolumn{2}{c}{\textbf{$m_\lambda$} \textbf{(mag)}}  & \textbf{$A_\lambda$} \textbf{(mag)}  \\
                  & & \textbf{Star A}   & \textbf{Star B}    &     \\
    \hline
   PS1 & $g$ & $>22.4$& $>22.4$ & 29.05 \\
  ZTF & $g$ & $>22.0$&  $>22.0$ & 30.25  \\ 
   PS1 & $r$ & $>22.4$&$>22.4$  & 20.80 \\
  ZTF & $r$ & $>22.5$&$>22.5$ & 20.92  \\ 
   PS1 & $i$ & $>22.3$& $>22.3$& 15.41 \\
   ZTF & $i$ & $>20.9$& $>20.9$ & 15.55 \\ 
   LDT & $i$ & $>23.7$& $>23.7$ & 15.55 \\
   PS1 & $z$ & $>21.5$& $>21.5$ & 12.11 \\
  LDT & $z$ & $>22.1$& $>22.1$& 11.57 \\
   PS1 & $y$ & $>20.5$& $>20.5$ & 9.96 \\
   UKIDSS & $J$ & $21.3\pm0.4$ & $18.81\pm0.05$ & 6.49 \\
   UKIDSS & $H$ & $18.62\pm0.07$ & $17.35\pm0.08$ & 4.11 \\
   UKIDSS & $K$ & $16.93\pm0.03$ & $17.35\pm0.03$ & 2.77 \\
   GLIMPSE & $3.6$\,$\mu$m & $15.72\pm0.07$& ... &  1.63 \\
   GLIMPSE & $4.5$\,$\mu$m & $15.60\pm0.08$& ... & 1.35 \\
   GLIMPSE & $5.8$\,$\mu$m & $15.62\pm0.15$& ... & 1.19 \\
    \hline
    \end{tabular}
\end{table}

\section{Results} 
\label{sec: results}

\subsection{Timing Analysis}
\label{sec: timing}

\subsubsection{Orbital period}
\label{sec: orbit_timing}

 \begin{figure}
\centering
\includegraphics[width=\columnwidth]{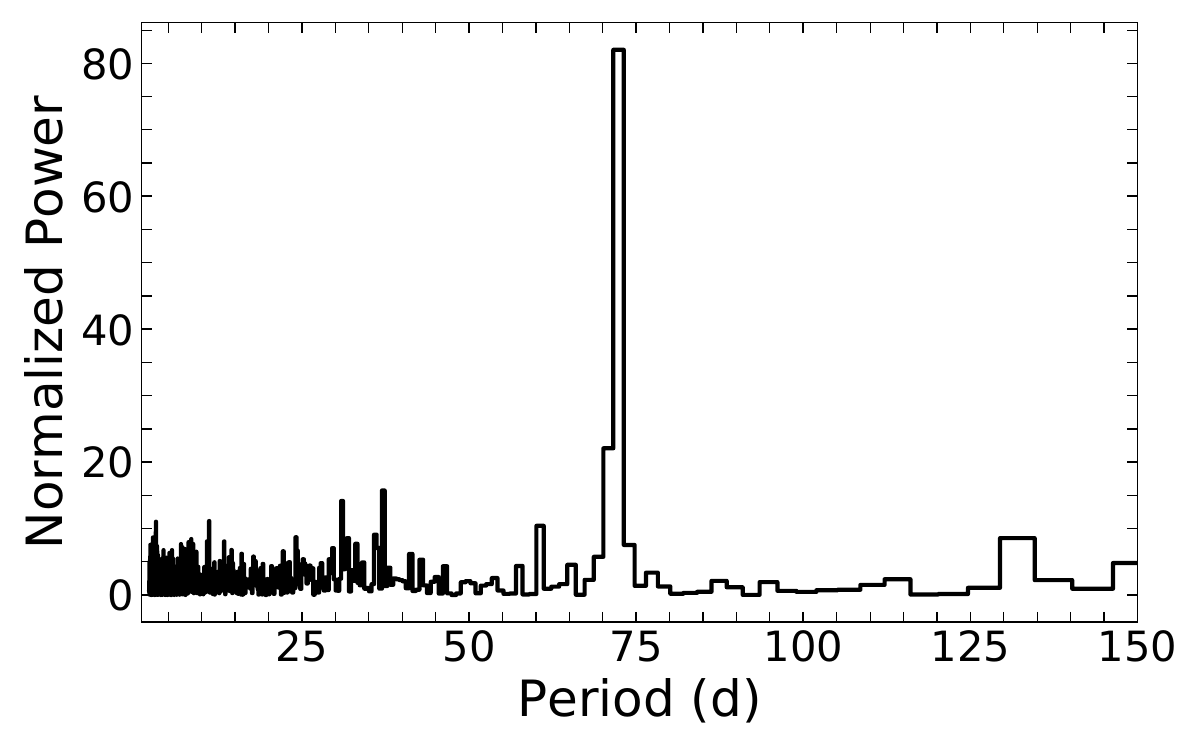}
\caption{Lomb-Scargle periodogram of the 9.2\,yr of \textit{Swift}/BAT monitoring. The peak corresponds to an orbital period at $P_\textrm{orb}=72.3\pm 0.3$ d.
}
\label{fig: BAT_LS}
\end{figure}

\begin{figure*} 
\centering
\includegraphics[width=1.5\columnwidth]{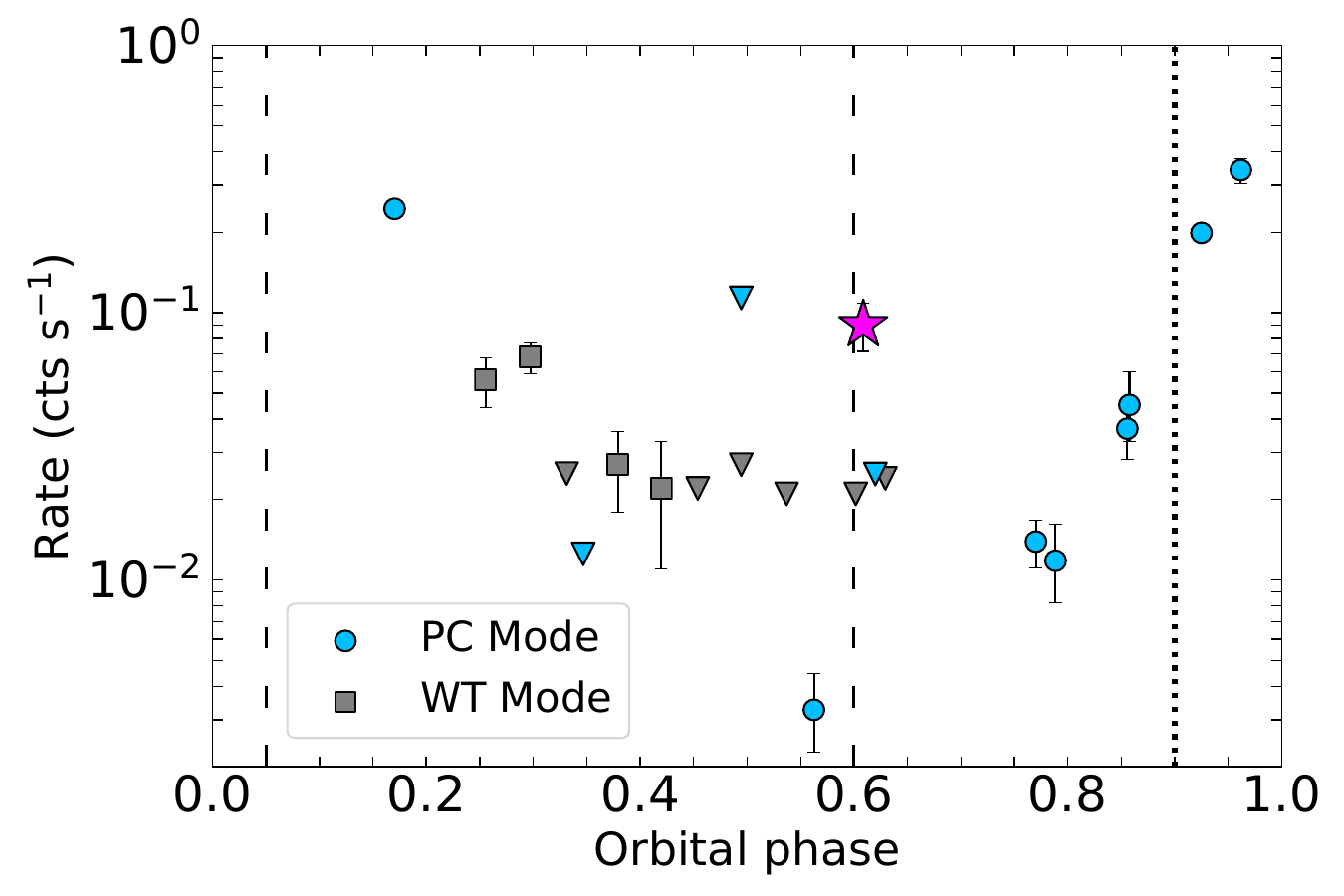} 
\caption{\textit{Swift}/XRT observations of J18219 as a function of orbital phase. Data in PC (WT) mode are shown with blue circles (gray squares); $3\sigma$ upper limits are denoted by downward triangles. The magenta star (PC mode data) represents a significant outlier - an outburst occurring close to apastron on MJD 58177. Dashed (dotted) vertical lines represent the orbital phase of our \textit{NICER} (\textit{NuSTAR}) observations.}
\label{fig: swift_orb_phase}
\end{figure*}

We used the long-term \textit{Swift}/BAT monitoring data (see \S \ref{sec: BAT}) to search for a periodic signal, as previously reported by \citet{LaParola2013}. A Lomb-Scargle frequency analysis \citep{Scargle1982} revealed an orbital period at $P_\textrm{orb}=72.3\pm 0.3$ d (see Figure \ref{fig: BAT_LS}), consistent with the period ($72.4\pm0.3$ d) derived by \citet{LaParola2013}. We calculate a false alarm probability of $2\times 10^{-31}$ \citep{Baluev2008}. Our analysis covers  $\sim$\,9.2 yr (corresponding to $\sim$\,$45$ orbits) of BAT observations confirming the orbital period of the system. Throughout this work, we define the orbital phase with respect to MJD 54656.26 assuming a period $P_\textrm{orb}\approx72.4$\,d for comparison with \citet{LaParola2013}. However, here we define the orbital phase of periastron passage as 0.0 phase, whereas in \citet{LaParola2013} the periastron passage (peak of BAT epoch folded lightcurve) occurs at 0.51 phase. In Table \ref{tab:Xobservations}, we report the orbital phase of all X-ray observations used in this work.


Figure \ref{fig: swift_orb_phase} displays the \textit{Swift}/XRT PC and WT mode observations as a function of orbital phase. We observe a clear trend of the source brightening and fading over the course of its orbit as it approaches and departs periastron passage. However, we also note the presence of a single X-ray detection occurring very close to apastron on MJD 58177, shown by the magenta star. 
At the same time, several observations at a similar orbital phase ($\sim$\,0.6) to the  magenta point resulted in upper limits, which leads us to conclude that such source behavior is uncommon. If such an outburst were to occur at apastron in every cycle,  based on our upper limits the source brightness would have to increase by a factor of $\gtrsim$\,$4$ within $0.5$ d, and decrease again by the same factor within $0.8$ d. 

To further explore this scenario, we observed the source with \textit{NICER} at phase $\sim$\, 0.6; the source was not detected with an upper limit to the unabsorbed flux of $<1.5\times 10^{-11}$ erg cm$^{-2}$ s$^{-1}$ (0.4-12 keV). Our \textit{NICER} (May 2021; Table \ref{tab:Xobservations}) observation would have been sensitive to an outburst similar to that observed on MJD 58177, which had an estimated unabsorbed flux $\sim$\,$3\times 10^{-11}$ erg cm$^{-2}$ s$^{-1}$ (0.3-10 keV), assuming the best fit model spectrum (see \S \ref{sec: avgspec} and Table \ref{tab: spectral_table}). Finally, a search in the daily BAT lightcurve ($15-50$ keV) around MJD 58177 did not reveal significant evidence for an outburst ($15-50$ keV). We briefly discuss this apastron X-ray flux excess in \S \ref{sec: discussion}.


\subsubsection{Spin period}
\label{sec: spin_timing}

\begin{figure} 
\centering
\includegraphics[width=\columnwidth]{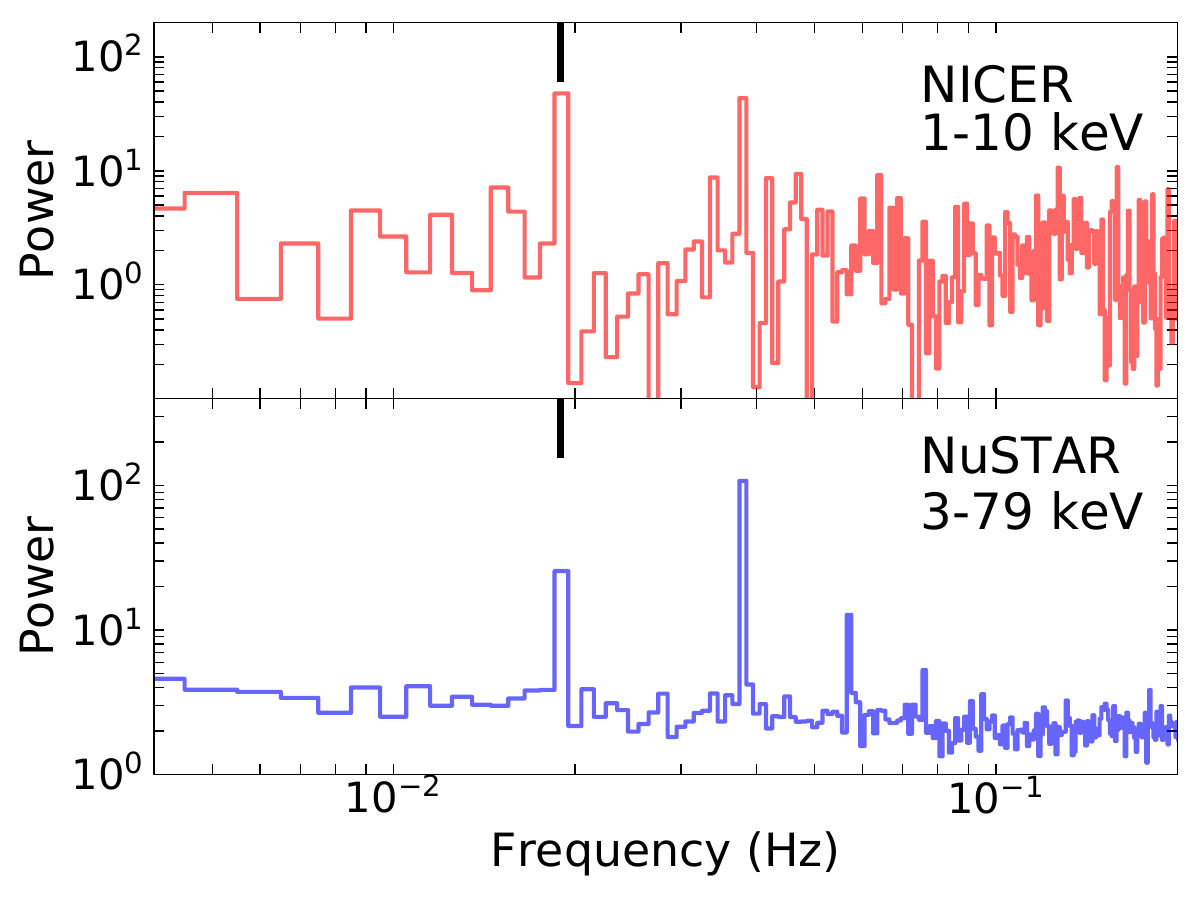}
\caption{Leahy normalized periodogram for our \textit{NICER} ($1-10$ keV; top) and \textit{NuSTAR} ($3-79$ keV; bottom) observations. The spin frequency (marked by black vertical line) is found at $0.019$ Hz, corresponding to a period of $52.46$ s. A number of harmonics of this frequency are also observed at 0.038 Hz, 0.057 Hz, and 0.076 Hz.  
}
\label{fig: periodogram_spin}
\end{figure}

We searched the \textit{NuSTAR} (March 2020) and \textit{NICER} (June 2020; Table \ref{tab:Xobservations}) data for coherent pulsations using both Z$^{2}$ statistics \citep{Buccheri1983} and by building the Leahy normalized power spectral density (PSD). 
The Leahy normalized periodogram \citep{Leahy1983} for these observations, computed using \texttt{Stingray} \citep{Huppenkothen2019}, is shown in Figure \ref{fig: periodogram_spin}. The periodogram was built from the \textit{NuSTAR} and \textit{NICER} lightcurves with events binned in time intervals of $\delta t=2\,$s, and averaged over segments with duration of $\tau=10^3\,$s. Therefore, the \textit{NuSTAR} periodogram is averaged over 46 segments (including both FPMA/B), whereas \textit{NICER} is averaged over only 2 segments due to the shorter exposure, using the \texttt{AveragedPowerspectrum} task within \texttt{Stingray}. 

We identify strong pulsations in \textit{NuSTAR} at the frequency of 0.0190593(1) Hz. This corresponds to a period $P_\textrm{spin}=52.4680\pm0.0003$ s, which we interpret as the spin period of a NS in the binary system. \textit{NICER} observations $\sim80$ days later also show a coherent signal at 0.01906(1) Hz, yielding a period of $52.466\pm0.007$ s.
The two spin frequencies are consistent with each other within $1\sigma$. 


In addition, we observe a number of harmonics of the spin frequency in the PSD at $0.038$ Hz in \textit{NICER}, and $0.038$, $0.057$, and $0.076$ Hz in \textit{NuSTAR} (Figure \ref{fig: periodogram_spin}), corresponding to $n=2$, $3$, and $4$ in the Fourier series decomposition, which are all detected at a $>3\sigma$ CL.  We note that the most significant peak in the \textit{NuSTAR} Leahy normalized periodogram is located at 0.0381185(2) Hz, but we disregard this as the fundamental frequency due to the presence of the peak at $0.057$ Hz, which is not an expected harmonic of $0.038$ Hz. We also note that using Z$^2_n$ epoch-folding statistics \citep{Buccheri1983}, where $n$ is the number of harmonics, leads to a higher significance peak at $0.019$ Hz when $n=2$ and $3$ (i.e., Z$^2_3$) as the pulse profile is not strictly sinusoidal (i.e., $n$\,$=$\,$1$, which favors $0.038$ Hz).

We additionally searched for a similar timing feature in our \textit{Swift}/XRT PC mode data and our \textit{NICER} observation from May 2021, but due to the low number of counts we were unable to find a significant peak at the expected frequency. 


\begin{figure} 
\centering
\vspace{-15mm}
\includegraphics[width=\columnwidth]{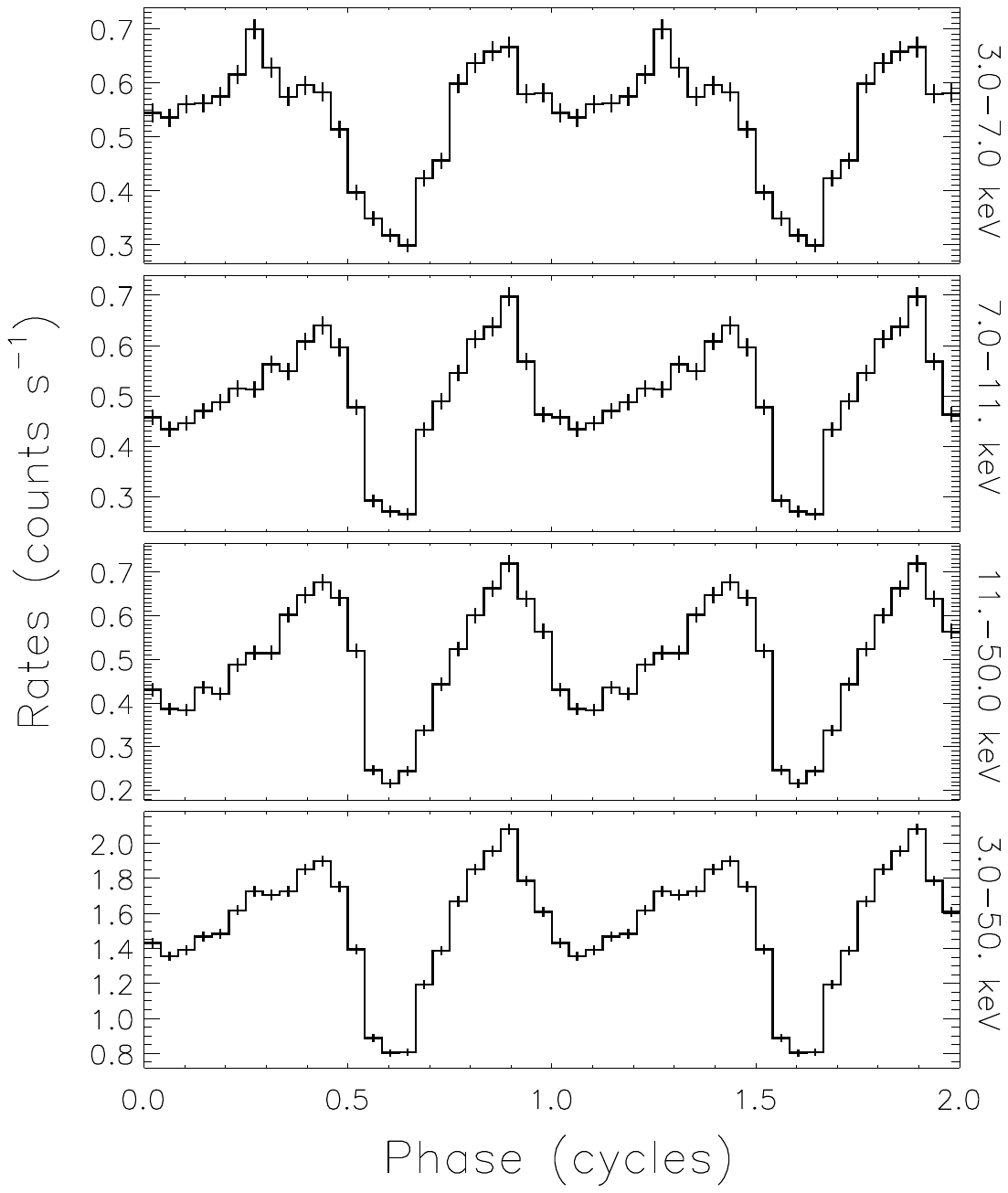}
\vspace{-15mm}
\caption{Phase-folded lightcurve from our \textit{NuSTAR} (FPMA/B) observation.
}
\label{fig: phase_profile_nustar}
\end{figure}

\begin{figure} 
\centering
\vspace{-15mm}
\includegraphics[width=\columnwidth]{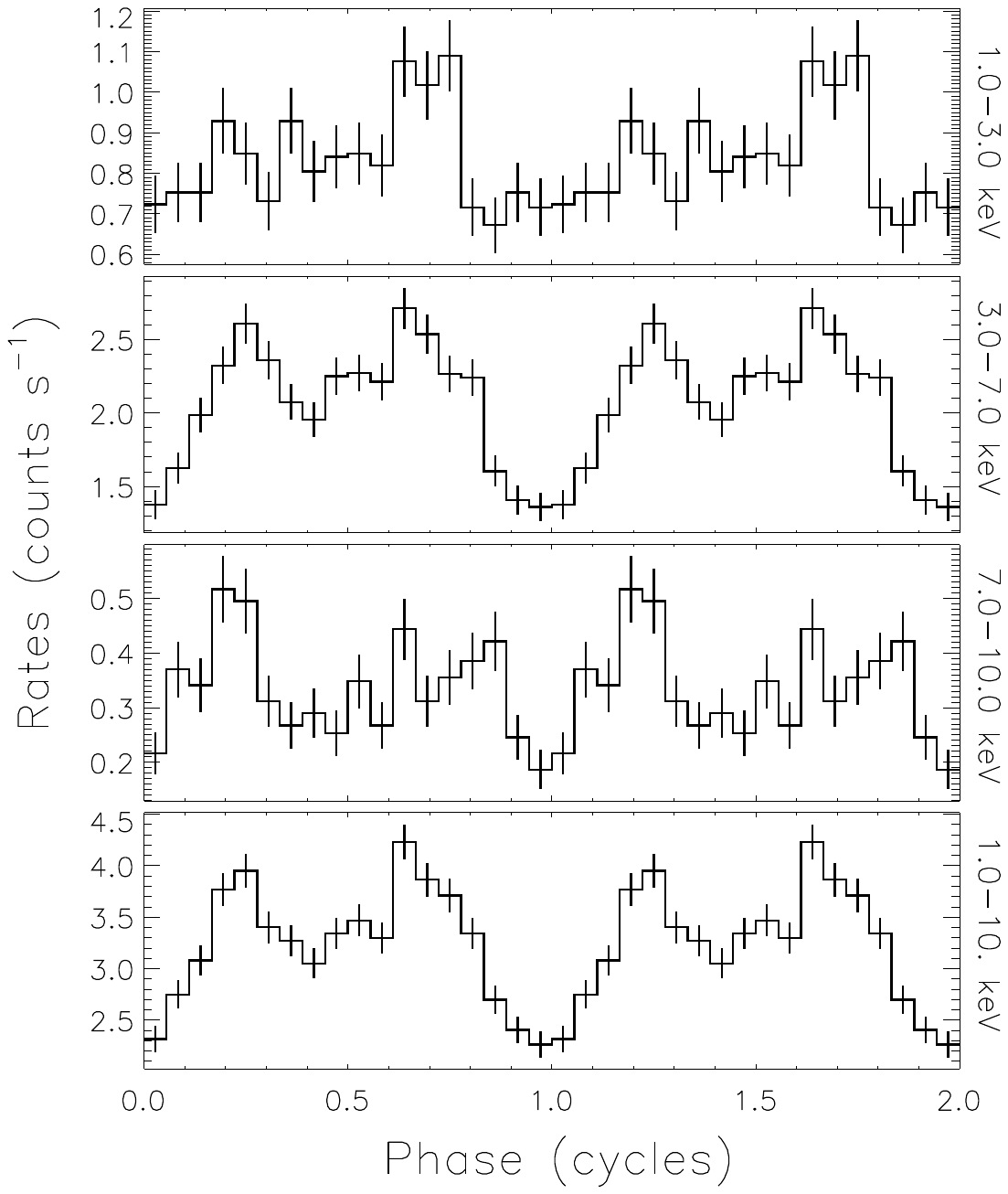}
\vspace{-15mm}
\caption{Phase-folded lightcurve from our \textit{NICER} observation on June 2020. 
}
\label{fig: phase_profile_nicer}
\end{figure}

In Figures \ref{fig: phase_profile_nustar} and \ref{fig: phase_profile_nicer} we present the pulse profiles for \textit{NuSTAR} and \textit{NICER} (June 2020 observation) in several energy bands. In \textit{NuSTAR} a well defined pulse profile is detected in all energy bands, whereas in \textit{NICER} the pulsations are very weak in the $1-3$ keV band (due to the high {\it NICER} background below $1.5$ keV), but increase in strength above $3$ keV. The phase-folded profiles display a similar shape at the common energy bands (i.e, $3-7$ keV) in \textit{NuSTAR} and \textit{NICER}. 
The two peaks in both phase-folded lightcurves are separated by $\sim$\,$0.5$ phase, which in combination with their similar peak heights drives the appearance of the harmonics in the periodogram. 

\begin{figure} 
\centering
\includegraphics[width=\columnwidth]{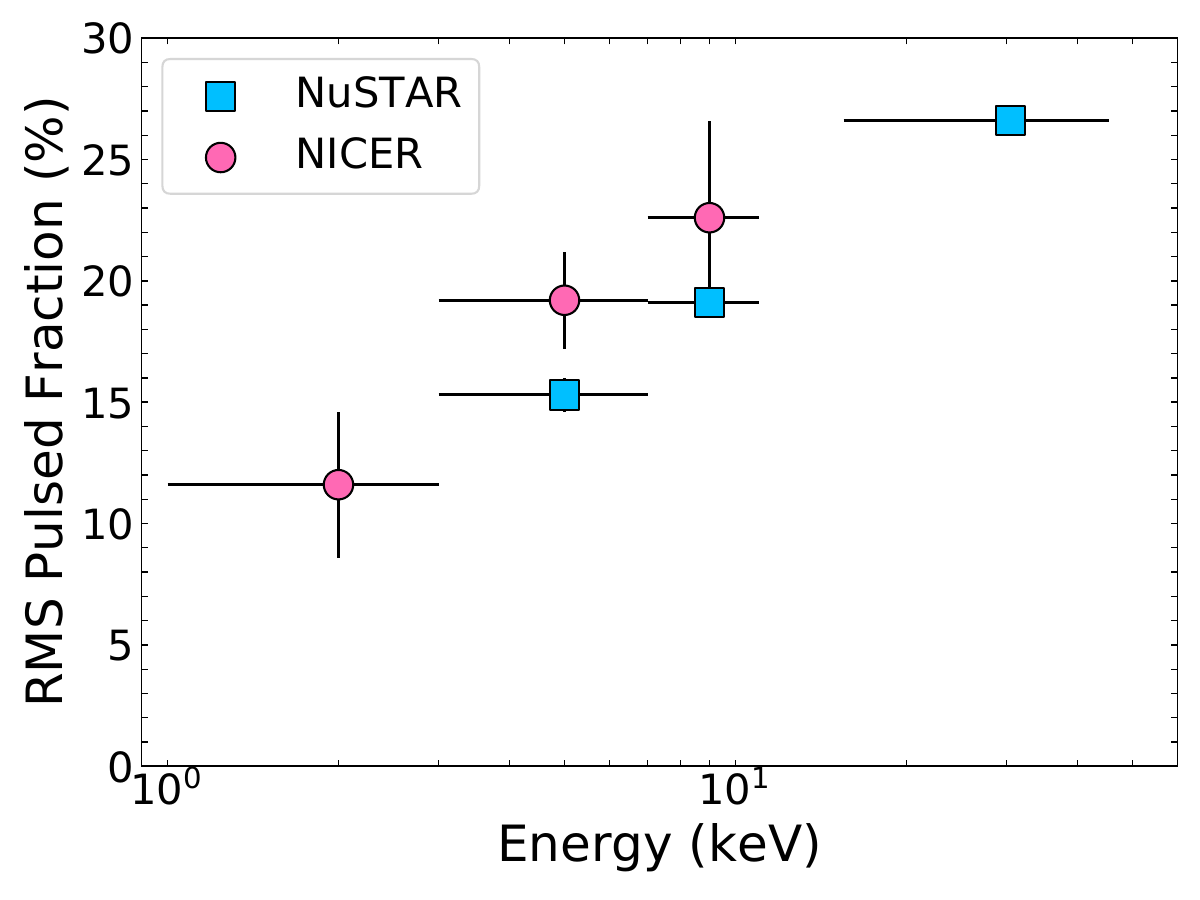}
\caption{RMS pulsed fraction as a function of energy for our \textit{NuSTAR} and \textit{NICER} (June 2020) observations.}
\label{fig: rmsfraction}
\end{figure}

We computed the root mean square (RMS) pulsed fraction in these energy bands for both instruments using the definition from \citet{Dhillon2009} (their equation 2). We observe a clear trend in the RMS pulsed fraction: it increases from $\sim20\%$ in the soft ($3-7$\,keV) band to $\sim28\%$ in the harder ($11-50$\,keV) band as displayed in Figure \ref{fig: rmsfraction}. This trend of increasing pulsed fraction with energy is commonly observed in X-ray pulsars within HMXB systems \citep{Lutovinov2008}.
The pulsed fraction in the full band for each instrument is $22.0\pm0.3$\% for \textit{NuSTAR} ($3-50$\,keV) and $16.5\pm1$\% for \textit{NICER} ($1-10$\,keV).

\subsubsection{\textit{NuSTAR} Lightcurve Variability}
\label{sec: lightcurve}

\begin{figure*} 
\centering
\includegraphics[width=2.0\columnwidth]{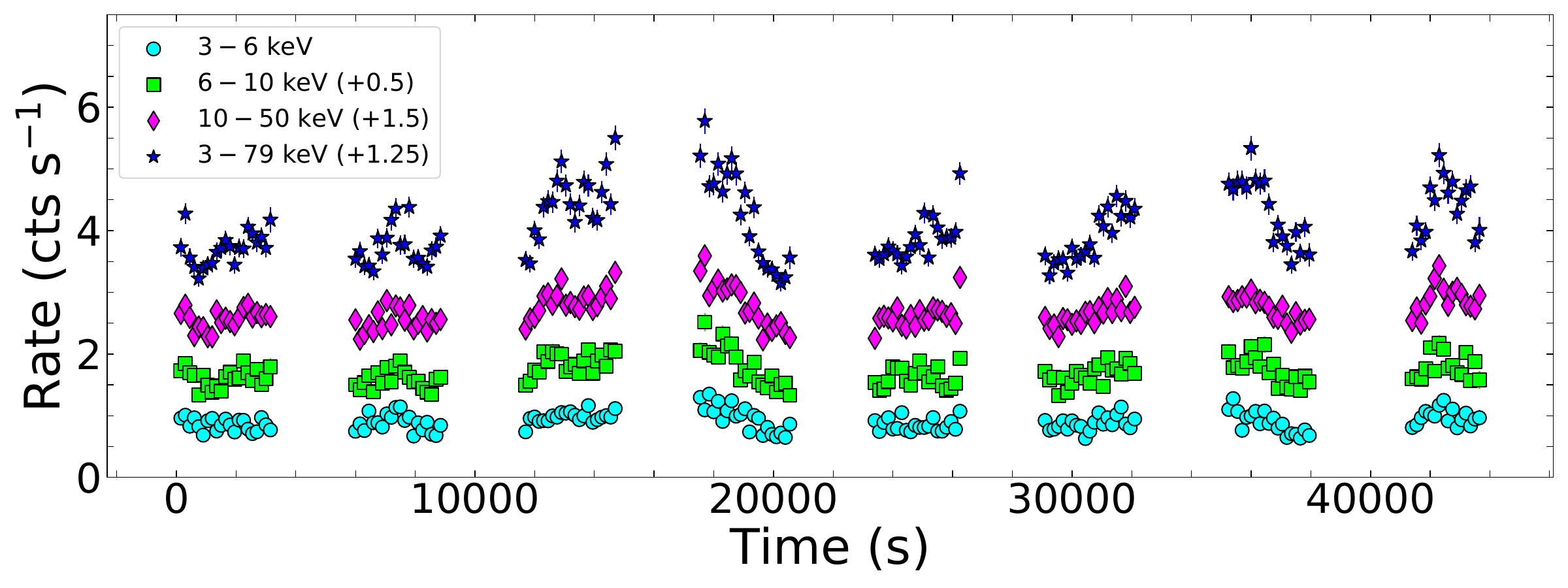}
\caption{\textit{NuSTAR}/FPMA lightcurve of J18219 in the $3-6$, $6-10$, $10-50$, and $3-79$ keV energy range with a time bin of 110 s. The $6-10$, $10-50$, and $3-79$ keV lightcurves have been shifted upwards by $0.5$, $1.5$, and $1.25$ cts s$^{-1}$, respectively.
}
\label{fig: nustar3-79kev}
\end{figure*}

In Figure \ref{fig: nustar3-79kev}, we show the \textit{NuSTAR}/FPMA lightcurve in the $3-6$ keV, $6-10$ keV, and $10-50$ keV energy ranges. The lightcurve displays variability on a timescale of $\sim$\, $5000-6000$ s. To explore whether this is identified as a timing feature, we again used \texttt{Stingray} to build the Leahy normalized periodogram, instead using the \texttt{Powerspectrum} task with the lightcurve binned in time intervals of $\delta t=100\,$s. The orbital gaps in the lightcurve were filled with white noise to minimize the effect of \textit{NuSTAR}'s low-Earth orbit duration ($\sim 5800$\,s) on the periodogram. This analysis was performed for both the FPMA and FPMB lightcurves individually, as well as the combined FPMA/B lightcurve.
We did not identify either a coherent or quasi-periodic oscillation on the timescale of the observed lightcurve variability ($\sim$\, $5000-6000$ s).
In fact, the power spectrum was found to be consistent with stochastic (red) noise \citep{Press1978}.

The lightcurve variability is visible across all energies ($3-6$ keV, $6-10$ keV, and $10-50$ keV) with a consistent trend between the different energy bands (Figure \ref{fig: nustar3-79kev}). To probe the nature of this variability, we further explore spectral variability of the source over these timescales in \S \ref{sec: spec_lightcurve_modulations}.

\subsection{Spectral Analysis}
\label{sec: specanalysis}

\subsubsection{Time-averaged spectroscopy}
\label{sec: avgspec}


We performed a time-averaged spectral analysis of the \textit{NuSTAR} observation in the $3-50$ keV energy band using \texttt{XSPEC} v12.11.0 \citep{Arnaud1996}. Both the FPMA and FPMB spectra were fit simultaneously with a prefactor. The normalization of FPMA $N_\textrm{FPMA}$ was fixed to unity and we allowed the normalization of FPMB $N_\textrm{FPMB}$ to vary. The prefactor $N_\textrm{FPMB}$ varied by $\lesssim10\%$ compared to unity, likely due to a rip in the multi-layer insulation of FPMB \citep[see][]{Madsen+2020}. We fit the spectra with an absorbed cutoff power-law (model \texttt{con*tbabs*cutoffpl}) with the ISM abundance table set using the command \texttt{abund wilm} \citep{Wilms2000}. This resulted in a good  spectral fit \citep[Cstat = 1377 for 1245 degrees of freedom, hereafter dof;][]{Cash1979}. We then used the \texttt{cflux} model to derive the time-averaged unabsorbed flux for the model. 

Finally, we tested an absorbed power-law model, which provided a much worse fit of the data (Cstat = 2154 for 1246 dof). We, therefore, consider the absorbed cutoff power-law to be the best fit model for the time-averaged flux and report its parameter values in Table \ref{tab: spectral_table}.




\begin{figure}
\centering
\includegraphics[width=1.\columnwidth]{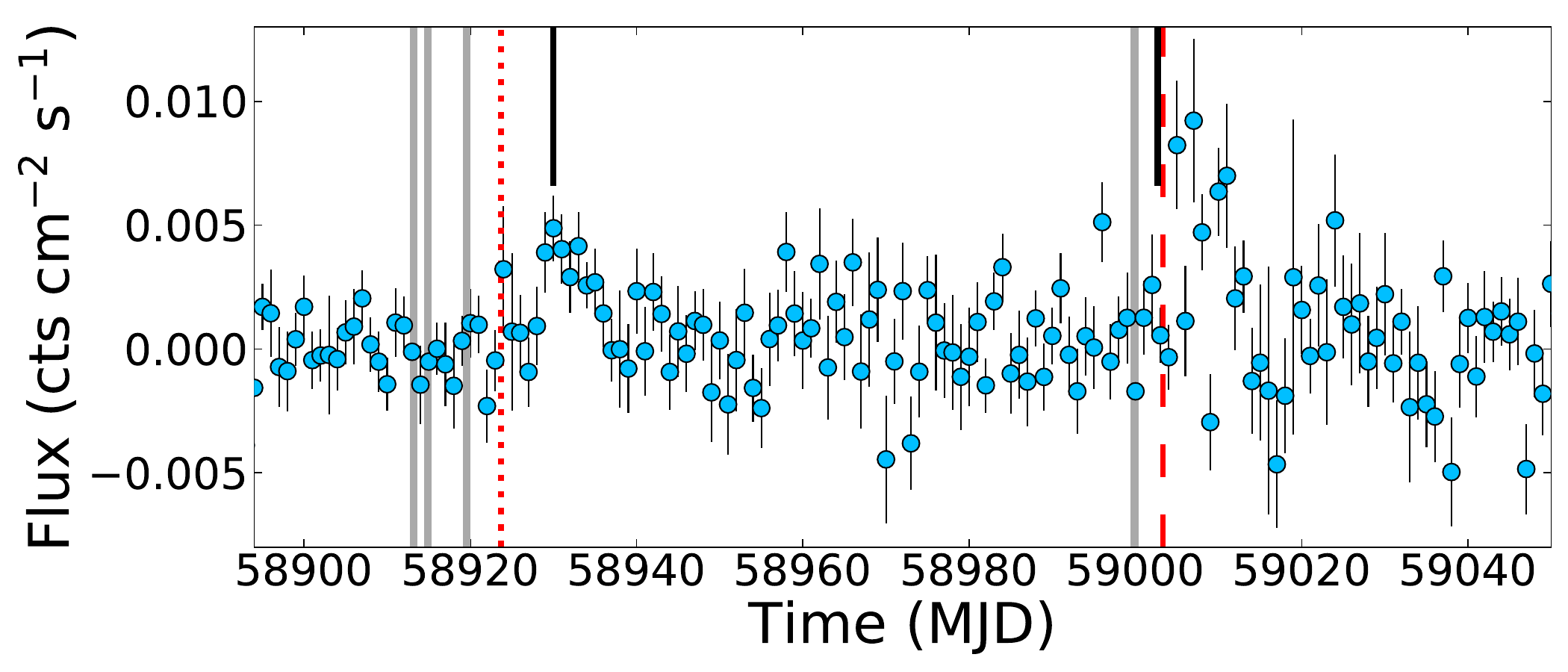} 
\caption{\textit{Swift}/BAT lightcurve including our \textit{NuSTAR} and \textit{NICER} observations in March and June 2020 (dotted and dashed red lines, respectively). Solid black lines mark the expected periastron passage of the NS, and solid gray lines mark our \textit{Swift}/XRT observations.
}
\label{fig: BAT_lightcurve}
\end{figure} 

We next fit the spectrum of our \textit{NICER} observations also with an absorbed cutoff power-law ($1.5-10$\,keV). 
Due to the narrower spectral range of  \textit{NICER}, the cutoff power-law model did not provide a meaningful constraint on the cutoff energy. Therefore, we fixed the cutoff energy to $14$ keV, in agreement with the value derived in our \textit{NuSTAR} spectrum ($14.7\pm0.6$ keV). The results of this analysis are also presented in Table \ref{tab: spectral_table}.
 
\begin{figure} 
\centering
\includegraphics[width=\columnwidth]{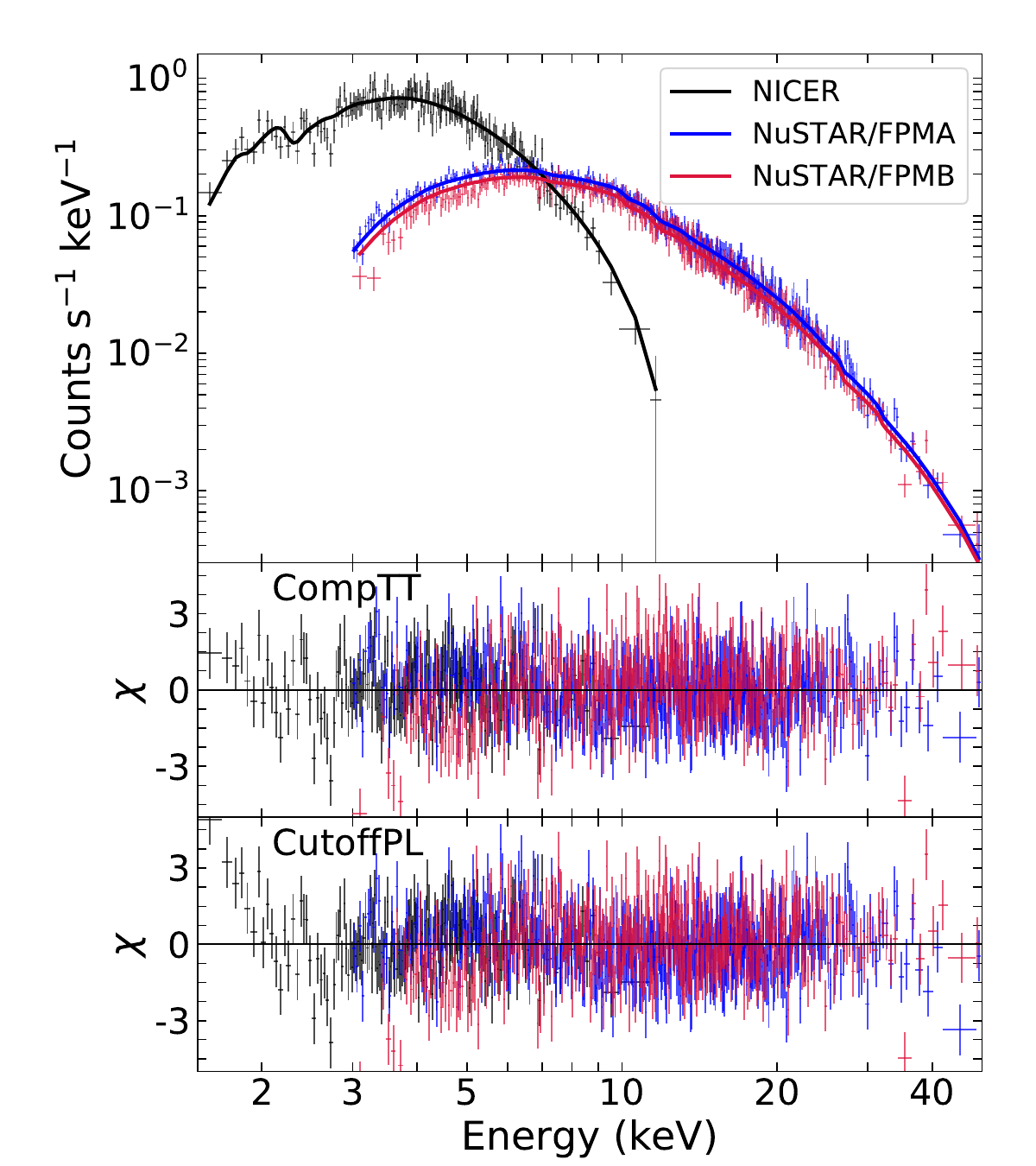}
\caption{Joint \textit{NuSTAR} and \textit{NICER} fit ($1.5-50$\,keV). The combined spectrum is well described by a thermally Comptonized model (top panel). The middle panel displays the fit residuals for the \texttt{CompTT} model, whereas the bottom panel shows the residuals for the absorbed cutoff power-law model.
}
\label{fig: joint_nicer+nustar_fit}
\end{figure}

Our \textit{NuSTAR} and \textit{NICER} observations were obtained in a similar orbital phase: \textit{NuSTAR} at phase $\sim$\,0.9 and \textit{NICER} at phase $\sim$\,0.01 (see Figure \ref{fig: BAT_lightcurve}). We expected, therefore, that the NS interaction with the Be decretion disk would be similar in both observations. We performed, therefore, an additional joint fit, including both \textit{NuSTAR} and \textit{NICER} spectra, using an absorbed cutoff power-law model. We allowed the normalization of the \textit{NICER} spectrum to vary with respect to the normalization of \textit{NuSTAR}/FPMA, yielding a value of $\sim$\,$0.87$.
The results of this analysis are included in Table \ref{tab: spectral_table}, and the fit residuals are displayed in the bottom panel of Figure \ref{fig: joint_nicer+nustar_fit}. 
In addition we tested a joint fit including the combined \textit{NuSTAR},~ \textit{NICER}, and \textit{Swift}/XRT PC mode spectra, and  obtained the same result as in Table \ref{tab: spectral_table} (\textit{NuSTAR} \& \textit{NICER} column) with no variation in the fit parameters or their errors.

 Our time-averaged spectral results are consistent with the combined \textit{Swift}/XRT and \textit{Swift}/BAT spectral analysis presented in \citet{LaParola2013}, albeit with smaller uncertainty on the fit parameters. We do note, however, that the $N_H$ inferred by \citet{LaParola2013} of $4.3^{+3.8}_{-1.7}\times10^{22}$ cm$^{-2}$ is smaller, but consistent at the $2\sigma$ level with our value. Both $N_H$ values are in excess of the Galactic value, $N_{H,\textrm{gal}}=1.49\times 10^{22}$ cm$^{-2}$ \citep{Willingale2013}, implying a potentially significant contribution intrinsic to the source environment. We tested whether this excess $N_H$ was required by performing a joint \textit{NuSTAR} and \textit{NICER} fit with fixed $N_H=1.49\times 10^{22}$ cm$^{-2}$. This resulted in a very poor fit to the data (Cstat = 3478 for 1929 dof) with significant residuals compared to the best fit model (Cstat = 2130 for 1929 dof).

The phenomenological cutoff power-law model suggests a physical emission mechanism of thermal inverse Compton scattering \citep{Titarchuk1994}. We, therefore, fit the broadband X-ray spectrum (\textit{NuSTAR} and \textit{NICER}) with a thermally Comptonized model \texttt{CompTT} in \texttt{XSPEC} (\texttt{con*tbabs*CompTT}; \citealt{Titarchuk1994}). The analytical \texttt{CompTT} model is described by the temperature of soft X-ray seed photons of temperature $kT_0$, which are Comptonized by a hot plasma with temperature $kT_1$ and optical depth $\tau$. We find that this model provides an improved description of the low energy ($<2$ keV) emission observed in \textit{NICER} with smaller residuals (Cstat = 2079 for 1928 dof; see Figure \ref{fig: joint_nicer+nustar_fit}). The best fit model (Table \ref{tab: comptt}) has $kT_0=1.36\pm0.03$ keV and $kT_1=6.98\pm0.13$ keV with optical depth $\tau=5.19\pm0.11$. The fit also resulted in a smaller Hydrogen column density of $N_H=(4.79\pm0.25)\times 10^{22}$ cm$^{-2}$ compared to that implied by the phenomenological cutoff power-law model.

\begin{table*}
    \centering
    \caption{Time-averaged and time-resolved spectral analysis results of J18219 X-ray observations using an absorbed cutoff power-law (\texttt{tbabs*cutoffpl}). 
    }
    \label{tab: spectral_table}
    \begin{tabular}{lcccc}
    \hline
    \hline
  \textbf{Time-averaged} &  \textit{NICER}$^a$ & \textit{NuSTAR}  & \textit{NuSTAR} \& \textit{NICER} \\ 
    \hline
    \hline
           $N_H$ ($10^{22}$ cm$^{-2}$) & $7.4\pm0.4$ & $11.2\pm0.8$ & $8.3\pm0.3$   \\
      $\Gamma$  & $0.37\pm0.07$ & $0.65\pm0.04$ &  $0.51\pm0.03$   \\
    $E_\textrm{cut}$ (keV)   & 14 (frozen) & $14.7\pm0.6$ &  $13.3\pm0.04$   \\
    $N_\textrm{FPMA}$ & -- & 1.0 & 1.0 \\
    $N_\textrm{FPMB}$ & -- & $0.97\pm0.01$ & $0.94\pm0.02$ \\
    $N_\textit{NICER}$ &1.0 & -- & $0.87\pm0.02$\\
       $F_X^b$ ($10^{-10}$ erg cm$^{-2}$ s$^{-1}$)  & $0.67\pm0.012$ & $1.93\pm0.02$ &  $1.37\pm0.02$    \\
      Cstat & 734 (685 dof) & 1377 (1245 dof) &  2130 (1929 dof)   \\
   \hline
     \hline
       \textbf{\textit{NuSTAR} time-resolved} & \textit{Decreasing}  & \textit{Increasing} & \textit{Linked $N_H$} \& \textit{$E_{cut}$} \\ 
    \hline
    \hline
  $N_H$ ($10^{22}$ cm$^{-2}$)  & $10.1\pm1.8$  & $12.5\pm1.9$   & $10.4\pm1.3$ \\
  $\Gamma$  & $0.70\pm0.10$ &$0.72\pm0.10$ &  $0.67\pm0.07$ \\
  $\Gamma_2^c$  & -- & --  & $0.70\pm0.07$ \\
  $E_\textrm{cut}$ (keV)  & $16.3\pm1.7$  & $16.9\pm1.9$ & $16.7\pm1.3$  \\
  $N_\textrm{FPMA}$ & 1.0 & 1.0 & 1.0 \\
    $N_\textrm{FPMB}$ & $0.94\pm0.01$  & $0.95\pm0.01$ & $0.94\pm0.01$ \\
  $F_X$ ($10^{-10}$ erg cm$^{-2}$ s$^{-1}$)  & $2.12\pm0.07$  & $1.93\pm0.07$  & $1.98\pm0.04$ \\
  $F_{X,2}^c$ ($10^{-10}$ erg cm$^{-2}$ s$^{-1}$)  & -- & -- & $2.25\pm0.06$ \\
   Cstat & 652 (682 dof) & 604 (653 dof) & 1330 (1335 dof)  \\
        \hline
     \hline
    \end{tabular}
    \begin{flushleft}
     \footnotesize{$^a$ The \textit{NICER} only model flux is provided in the $2-10$ keV energy range.\\
     $^b$ Unabsorbed flux ($3-50$\,keV). \\
     $^e$ Decreasing state only. \\
     }
\end{flushleft}
\end{table*}

\begin{table}
    \centering
    \caption{Results for a joint \textit{NuSTAR} and \textit{NICER} spectral fit with a thermally Comptonized (\texttt{CompTT}) model.
    }
    \label{tab: comptt}
    \begin{tabular}{lccc}
    \hline
    \hline
   \textbf{Parameter}  & \textbf{Value} & \textbf{Units}  \\
    \hline
    \hline
           $N_H$   & $4.79\pm0.25$  & $10^{22}$ cm$^{-2}$ \\
      $kT_0$ & $1.36\pm0.03$ & keV \\
      $kT_1$ & $6.98\pm0.13$ & keV \\
      $\tau$ & $5.19\pm0.11$ & \\
        $N_\textrm{FPMA}$ & 1.0 & \\
        $N_\textrm{FPMB}$ & $0.88\pm0.01$ & \\
        $N_\textit{NICER}$ & $0.86\pm0.02$ & \\
       $F_X$($3-50$ keV)  & $1.26\pm0.04$ & $10^{-10}$ erg cm$^{-2}$ s$^{-1}$   \\
      Cstat & 2079 (1928 dof) &  \\
     \hline
     \hline
    \end{tabular}
\end{table}

\subsubsection{Time-resolved spectroscopy}
\label{sec: spec_lightcurve_modulations}

In this section, we investigate whether  spectral variability can explain the flux variability observed with \textit{NuSTAR} on a scale of a few thousand seconds (Figure \ref{fig: nustar3-79kev}). We split the \textit{NuSTAR} lightcurve (FPMA and FPMB) into two groups: intervals of the lightcurve which are either increasing or decreasing in count rate. 
These intervals were selected based on the $3-79$ keV lightcurve displayed in Figure \ref{fig: nustar3-79kev}. In the event that the lightcurve obtained over an individual \textit{NuSTAR} orbit displays variability (i.e, a switch from increasing to decreasing, or vice versa), the increasing and decreasing intervals were chosen to reflect this variability such that only increasing portions of the lightcurve are included in the increasing spectral analysis. We note that some small portions of the lightcurve are not strictly increasing or decreasing, and, therefore, these portions were ignored in our analysis. 
We used \texttt{XSELECT} to define Good Time Intervals (GTIs) and to extract the spectra, which were then binned to a minimum of 25 counts per bin; we used the Cash statistic within \texttt{XSPEC} for the model fitting. We modeled the spectra with the phenomenological absorbed cutoff power-law as outlined in \S \ref{sec: avgspec}. We chose to apply this model, as opposed to \texttt{CompTT}, due to its smaller number of fit parameters.

We find that the increasing and decreasing states can be described by the same spectrum (absorbed cutoff power-law), within $1\sigma$ errors. In order to more precisely determine the normalization and photon index, we linked the Hydrogen column density and cutoff energy within \texttt{XSPEC}, requiring those parameters to be identical for both spectra. The results of these analyses are presented in Table \ref{tab: spectral_table}. We conclude that spectral variability cannot explain the observed flux modulation.

\subsubsection{Phase-resolved spectroscopy}
\label{sec: phasespec}

We performed a phase-resolved spectral analysis with the \textit{NuSTAR} FPMA/B data to determine if there is spectral variability over the NS spin period. 
We selected the GTIs following the \textit{NuSTAR} phase folded lightcurve displayed in Figure \ref{fig: phase_profile_nustar}. Based on the double peaked pulse profile, we selected four spectral groups: \textit{i}) the shoulder of the small peak (phase $0.0-0.25$) \textit{ii}) the small peak between phase $0.25-0.5$, \textit{iii}) the valley between phase $0.5-0.75$, and, lastly, \textit{iv}) the main peak at $0.75-1.0$ phase. These spectra were modeled using an absorbed cutoff power-law as in the previous section.

The Hydrogen column density, photon index, and cutoff energy were consistent within the $1-2\sigma$ level among the four phase-resolved spectra; only the normalization of the power-law was different, as expected based on our selected GTIs. We also confirmed that the deviation between parameters when using a \texttt{CompTT} model was at the same level. Therefore, following the previous section, we froze the Hydrogen column density and cutoff energy among the four spectra in order to resolve any difference in photon index arising as a function of phase. These results are displayed in Table \ref{tab: phase_spec_table}. The fit statistic of the joint fit of the eight spectra (including both FPMA and FPMB data) is Cstat = 1935 for 1933 dof. 
We find a marginal indication of spectral variability between the two peaks ($\Gamma=0.56\pm0.04$) and the soft shoulder emission ($\Gamma=0.82\pm0.04$) between phase $0.0-0.25$ in Figure \ref{fig: phase_profile_nustar}. The deviation between the two photon indices is at the $\sim$\,$3\sigma$ level.



\begin{table*}
    \centering
    \caption{Phase-resolved spectral analysis results of our \textit{NuSTAR} (FPMA/B) observations. }
    \label{tab: phase_spec_table}
    \begin{tabular}{lccccc}
    \hline
    \hline
  \textbf{\textit{NuSTAR} Phase-resolved} &  \textit{Shoulder} & \textit{Small Peak}  & \textit{Valley} & \textit{Main Peak} \\ 
  \textit{Phase}$^b$ & $0.0-0.25$ & $0.25-0.5$ & $0.5-0.75$ & $0.75-1.0$ \\
    \hline
    \hline
           $N_H$ ($10^{22}$ cm$^{-2}$) & $5.8\pm0.5^a$ & -- & --& --   \\
      $\Gamma$  & $0.82\pm0.04$ & $0.56\pm0.04$ & $0.69\pm0.05$ & $0.56\pm0.05$ \\
    $E_\textrm{cut}$ (keV)   & $14.5\pm0.6^a$ & -- & -- & --   \\
    $N_\textrm{FPMA}$ & 1.0$^a$ & -- & -- & --  \\
    $N_\textrm{FPMB}$ & $1.08\pm0.01^a$ & -- & -- & --  \\
    $A_\textrm{norm}$ ($10^{-3}$) & $4.0\pm0.3$ & $2.7\pm0.2$ & $2.8\pm0.2$ & $2.2\pm0.2$ \\
     Cstat & 478 (469 dof) & 530 (560 dof) & 357 (347) & 569 (568 dof) \\
   \hline
     \hline
    \end{tabular}
   \begin{flushleft}
    \footnotesize{$^a$ The Hydrogen column density, cutoff energy, and the normalization of FPMA/B were fixed among the four phase-resolved spectra. \\
    $^b$ Phase selection is based on Figure \ref{fig: phase_profile_nustar}.
    }
   \end{flushleft}
\end{table*}

\begin{figure} 
\centering
\includegraphics[width=\columnwidth]{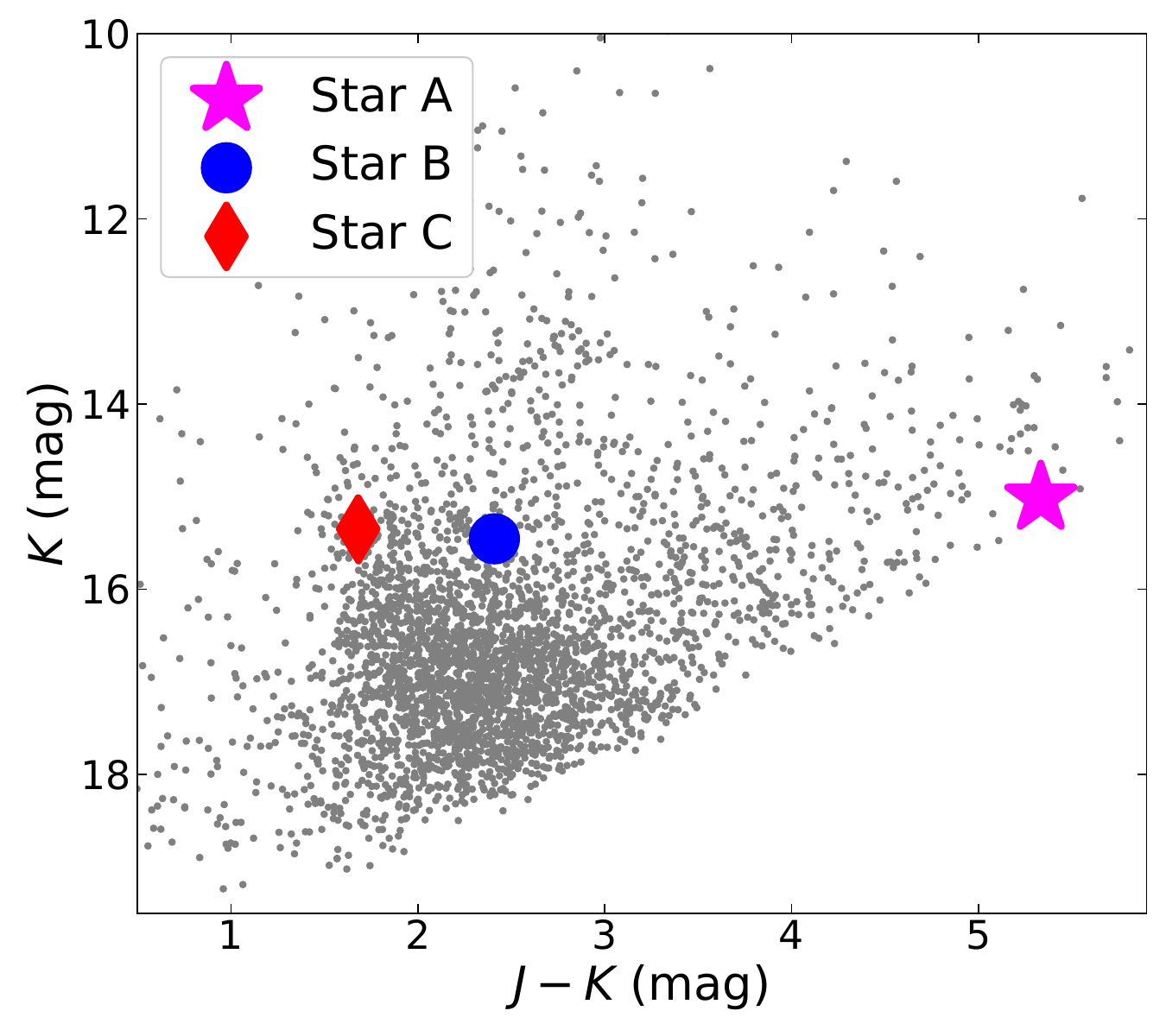}
\caption{Color-magnitude diagrams in the Vega magnitude system for the field of J18219 based on UKIDSS infrared imaging. Star A, B and C are represented by red, blue and green symbols, respectively. The magnitudes are not corrected for Galactic interstellar reddening.
}
\label{fig: CMDs}
\end{figure}

 \begin{figure*} 
\centering
\includegraphics[width=1.7\columnwidth]{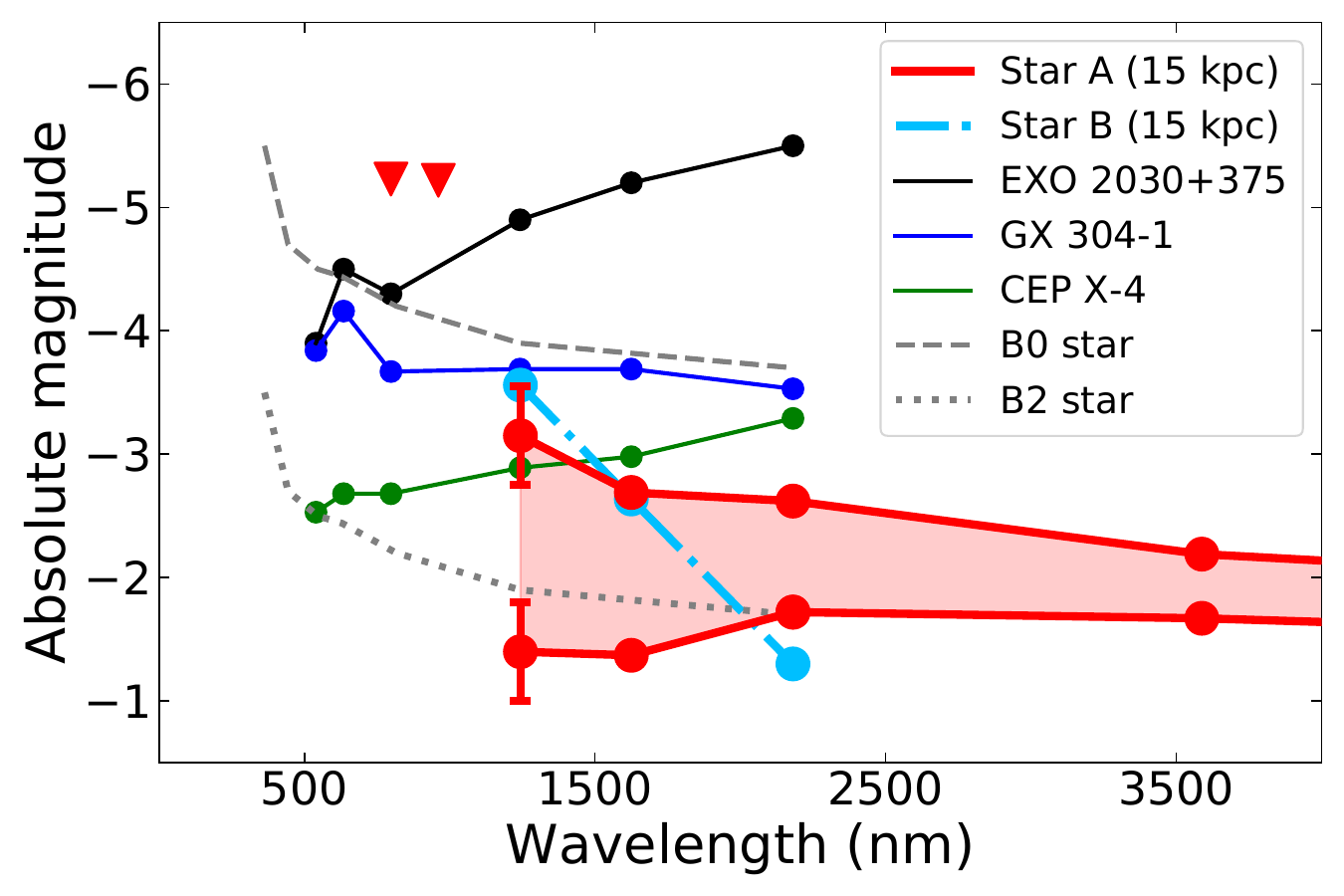}
\caption{Absolute (AB) magnitude, assuming a distance of 15 kpc, of Star A (red) and Star B (light blue) versus wavelength. Star A has been corrected for both Galactic extinction \citep[bottom red curve;][]{Schlafly2011} and the expected extinction assuming $N_H=8.3\times 10^{22}$ cm$^{-2}$ \citep[top red curve;][]{Guver2009}; Star B is only shown corrected for Galactic extinction. These stars are compared with the SEDs of known BeXRBs \citep[EXO 2030+375, GX 304-1, and CEP X-4;][]{Coe1997,Riquelme2012,Reig2014} which have been de-reddened, as well as template SEDs for B0 (gray dashed line) and B2 (gray dotted line) type stars. The downward red triangles represent $3\sigma$ upper limits derived from LDT and archival PS1 imaging (corrected for Galactic extinction); the limits apply for both Star A and Star B.}
\label{fig: sed_comparison}
\end{figure*}

\subsection{Optical/infrared counterpart}
\label{sec: counterpart}

Here, we report on our search for the counterpart of J18219 with LDT, which we supplemented with archival imaging from PS1, ZTF, UKIDSS, 
and the Galactic Legacy Infrared Mid-Plane Survey Extraordinaire \citep[GLIMPSE;][]{Benjamin2003}. At the \textit{Chandra} localization of J18219, we identify a bright infrared counterpart catalogued by the UKIDSS survey. The counterpart appears to be the blended combination of two point sources (Star A and Star B; Figure \ref{fig: opt_finding_chart}).
In order to resolve Star A and Star B, we used \texttt{DAOPHOT} PSF photometry to de-blend the sources, as outlined in \S \ref{sec: ukidss}. 
We include the archival photometry of the Star A and B source complex from the GLIMPSE catalog, using the Vega to AB magnitude conversion from \citet{Papovich2016}. Due to the large PSF of the \textit{Spitzer Space Telescope}, we cannot de-blend the photometry from GLIMPSE. However, based on the source SEDs for Star A and Star B, we assume the majority of the contribution at those wavelengths (3.6-5.8 $\mu$m) is coming from Star A. The photometry of both stars is tabulated in Table \ref{tab: optcounterpart}.

We did not detect an optical source coincident with the infrared counterpart in our LDT and SALT imaging, or in archival PS1 and ZTF images. The $3\sigma$ upper limits at the source position are provided in Table \ref{tab: optcounterpart}. The lack of optical source detection is not unexpected given the level of interstellar reddening, $E(B-V)=9.16$ mag \citep[or $A_V=28.4$ mag, assuming a ratio of total to selective extinction of $R_V=3.1$;][]{Rieke1985,Schlafly2011}, in the direction of the source.

Finally, we discuss here the de-blended magnitudes of Stars A and B (see \S \ref{sec: ukidss}). We show the $K,\ J-K$ color-magnitude diagrams (CMDs) for the observed field of view ($\approx$\,$2\arcmin$\,$\times$\,$2\arcmin$ ) in Figure \ref{fig: CMDs} in the Vega magnitude system: Star A, B and C are over-plotted with red, blue and green star symbols, respectively. It is interesting to note that Star A is one of the reddest objects in the field of view, with $H-K \gtrsim$ 2.0 mag and $J-K \gtrsim$ 5.0 mag in the Vega magnitude system (not corrected for extinction). On the other hand, stars B and C are quite blue objects, with $H-K \approx$ 0.5 mag (Vega) for both sources and $J-K \approx$ 1.5 mag for star C and $\approx$ 2.0 mag for star B. 

Next, we performed a comparison (using a similar methodology to \citealt{Lutovinov2016}) between the SEDs of Star A and Star B (Table \ref{tab: optcounterpart}) with the well-known BeXRBs EXO 2030+375, GX 304-1, and CEP X-4 \citep{Coe1997,Riquelme2012,Reig2014} and with template SEDs for B0 and B2 type stars (see Figure \ref{fig: sed_comparison}). 
We found that a distance of $10-15$ kpc is required in order for the absolute luminosity of the stars to be consistent with the expected range of values for a Be star and for a star with B0-B2 spectral class. Given this distance, and the source's Galactic coordinates ($l,b$\,$=$\,$17.32^\circ$, $0.13^\circ$), J18219 is likely located beyond the Galactic center, and possibly as far as the Outer Scutum$-$Centaurus Arm of our Galaxy \citep{Dame2011,Armentrout2017}. 

We find that Star A is consistent with the expected SED shape of the Be and B-type comparison stars ($J-K$\,$\approx$\,0 AB mag), whereas Star B is too blue in color. We note that due to the large uncertainty on the de-blended $J$-band magnitude ($J=21.3\pm0.4$ mag; not de-reddened) of Star A, the contribution at those wavelengths can be treated as an upper limit. For Star A, we have assumed two different extinction values: \textit{i}) a Galactic extinction \citep[see Table \ref{tab: optcounterpart};][]{Schlafly2011} yielding $J-K$\,$\approx$\,$0.3\pm0.4$ AB mag, and \textit{ii}) using the linear relation between hydrogen column density, $N_H$, and optical extinction, $A_V$, from \citet[][see their Equation 1]{Guver2009}. In the latter case, we assumed $N_H=8.3\times 10^{22}$ cm$^{-2}$ (Table \ref{tab: spectral_table}), which yields $A_V\sim 37.6$ mag\footnote{This conversion is computed assuming the average Galactic value of $R_V=3.1$ \citep{Savage1979,Rieke1985}, but in principle there is scatter in $R_V$ between $\sim$2 to 5.5 allowing for more freedom in converting between $N_H$ and $A_V$.}. This results in $J-K$\,$\approx$\,$-0.5\pm0.4$ AB mag. The shaded region in Figure \ref{fig: sed_comparison} represents the SED shape produced between these two different extinction scenarios. In either case, the SED of Star A remains consistent with a Be star. We note that assuming a smaller value of extinction, or $R_V<3.1$, would imply an even redder color for the source, but would require increasing the distance to extreme values ($>20$ kpc).

Furthermore, we find that Star C is likewise inconsistent with a Be or B-type star, due to its color ($J-K$\,$\approx$\,$-2$ AB mag) and the fact that it is significantly brighter in the optical compared to the infrared. Thus, Star C is a very unlikely companion to J18219. We conclude that Star A is the true counterpart to J18219, and that its Be star classification solidifies J18219 as a BeXRB.

\section{Discussion}
\label{sec: discussion}

\begin{figure} 
\centering
\includegraphics[width=\columnwidth]{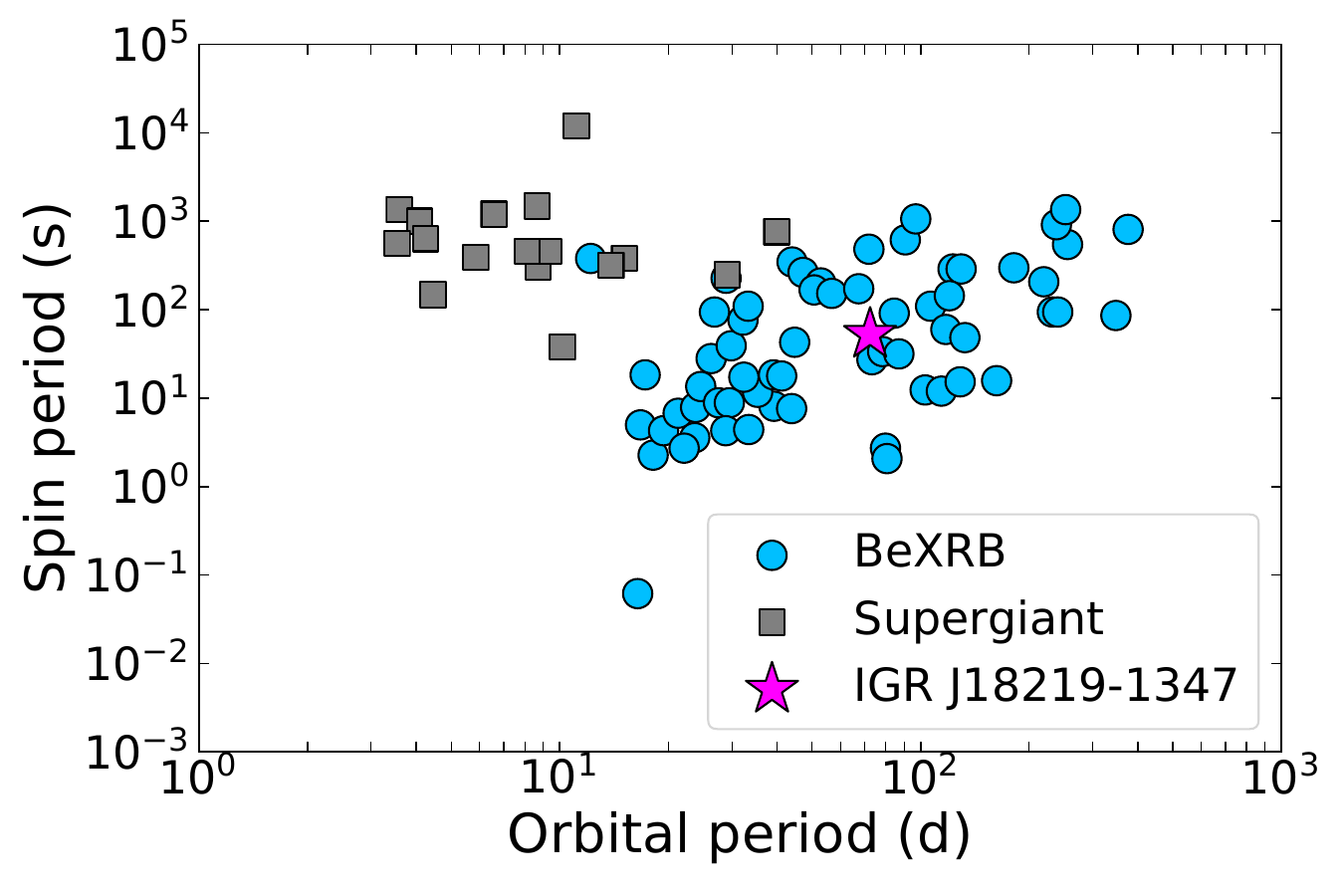}
\caption{Corbet diagram of HMXBs with a known orbital and spin period \citep{Liu2006,Corbet2017}. Gray squares mark the location of supergiant/X-ray binary systems, and blue circles represent known BeXRBs. The location of J18219 is marked by a magenta star. }
\label{fig: corbet}
\end{figure}

To further explore the nature of the binary system, we placed it in the Corbet Diagram \citep[Figure \ref{fig: corbet};][]{Corbet1986}; we found that it lies solidly in the region populated by known BeXRBs \citep{Liu2006,Corbet2017}. Additionally, it is located far from the population of supergiant/X-ray binaries (wind accreting systems) which generally exhibit shorter orbital periods and longer spin periods.
Thus, the determination of the NS spin period, $P_\textrm{spin}$, is vital information in the classification of the system. We argue, therefore, that the system's location near known BeXRBs, combined with the fact that the majority of its emission is over a small fraction of the orbit (Figure \ref{fig: swift_orb_phase}), indicates that the system is a BeXRB.

We used archival UKIDSS observations to de-blend the infrared counterpart into Stars A and B (Figure \ref{fig: opt_finding_chart}), resulting in the identification of Star A as a Be star (Figure \ref{fig: sed_comparison}).
In addition, the source SED (and X-ray luminosity), allowed us to place it at a distance between $10-15$ kpc. We note that at this distance the X-ray luminosity of the observed outburst by \textit{NuSTAR} is $(2-5)\times 10^{36}$ erg s$^{-1}$, which is toward the high end of the luminosity distribution of Type I outbursts in HMXBs \citep{Reig2011,Chaty2011}. 
We conclude that, combined with the X-ray properties, the counterpart's classification as a Be star is compelling and confirms the nature of J18219 as a BeXRB.

Finally, a similar detection to the possible apastron outburst from J18219 (\S \ref{sec: orbit_timing}) has been observed in only a handful of other BeXRBs \citep[e.g., EXO 2030+375;][]{Reig1998}. \citet{Reig1998} explained their  apastron outburst as originating from a Be star's wind with velocity equal to or smaller than the NS's orbital velocity, leading to efficient accretion onto the NS. Alternatively, such an outburst could ensue from a possible misalignment of the binary orbit with the Be star's disk. Unfortunately, the extremely limited archival X-ray data around this time period, do not allow further analysis as to the cause of this increase in brightness. Future monitoring of the source at apastron is required to discern whether such outbursts are regular, and uncover their nature.



\section{Conclusions}
\label{sec: conclusions}

We utilized \textit{Swift}, \textit{NuSTAR}, and \textit{NICER} observations to investigate the X-ray timing and spectral properties of J18219 in order to confirm the preliminary source classification as a BeXRB \citep{LaParola2013}. Through our timing analysis (\S \ref{sec: timing}), we uncovered a periodic signature in the \textit{NuSTAR} and \textit{NICER} lightcurves corresponding to a period $P_\textrm{spin}=52.46$ s. We interpret this as the spin period of a neutron star. Furthermore, using long term \textit{Swift}/BAT daily monitoring, we confirmed the orbital period of the system $P_\textrm{orb}=72.3\pm0.3$. Lastly, we confirmed that the infrared counterpart (Star A) is consistent with the expected SED of a Be star. 
These properties lead us to classify J18219 as a BeXRB.

We found that the time-averaged broadband X-ray spectrum ($1.5-50$ keV) obtained from \textit{NuSTAR} and \textit{NICER} was well described by either an absorbed cutoff power-law (\S \ref{sec: specanalysis}) with photon index $\Gamma\sim 0.5$ and cutoff energy $\sim 13$ keV or a thermally Comptonized model (Table \ref{tab: comptt}). The inferred Hydrogen column density from our spectral modeling (Table \ref{tab: spectral_table} and \ref{tab: comptt}) $N_H=(4-11)\times 10^{22}$ cm$^{-2}$ is well above the Galactic value of $N_{H,\textrm{gal}}=1.5\times 10^{22}$ cm$^{-2}$ \citep{Willingale2013}, requiring either a significant contribution from the environment of the binary system or an excess Galactic extinction in the line of sight, compared to the value implied by the 21 cm radio emission map of our Galaxy \citep[i.e., on a scale $<0.75$ degrees;][]{Kalberla2005,Willingale2013}. Future monitoring of the source over the course of its orbit will probe whether there is variability in the Hydrogen column density, shedding light on whether the contribution is intrinsic to the source.

\acknowledgments

The authors acknowledge useful discussions with Matteo Bachetti, Tomaso Belloni, and Oleg Kargaltsev. B.~O., C.~K., and N.~G. acknowledge supported under NASA Grants 80NSSC20K0389 and 80NSSC19K091. L.~J.~T. is supported by the South African National Research Foundation. J.~G. acknowledges support by the ISF-NSFC joint research program (grant no. 3296/19).

This work made use of data supplied by the UK \textit{Swift} Science Data Centre at the University of Leicester. This research has made use of the XRT Data Analysis Software (XRTDAS) developed under the responsibility of the ASI Science Data Center (ASDC), Italy. This research has made use of the NuSTAR Data Analysis Software (NuSTARDAS) jointly developed by the ASI Space Science Data Center (SSDC, Italy) and the California Institute of Technology (Caltech, USA). This research has made use of data and/or software provided by the High Energy Astrophysics Science Archive Research Center (HEASARC), which is a service of the Astrophysics Science Division at NASA/GSFC. The scientific results reported in this article are based on observations made by the Chandra X-ray Observatory. This research has made use of software provided by the Chandra X-ray Center (CXC) in the application package CIAO. These results also made use of Lowell Observatory's Lowell Discovery Telescope (LDT), formerly the Discovery Channel Telescope. Lowell operates the LDT in partnership with Boston University, Northern Arizona University, the University of Maryland, and the University of Toledo. Partial support of the LDT was provided by Discovery Communications. LMI was built by Lowell Observatory using funds from the National Science Foundation (AST-1005313). Some of the observations reported in this paper were obtained with the Southern African Large Telescope (SALT) under programme 2018-2-LSP-001. We additionally made use of Astropy, a community-developed core Python package for Astronomy \citep{astropy}.

%

\vspace{5mm}
\facilities{\textit{Swift}, \textit{NuSTAR}, \textit{NICER}, SALT, LDT, UKIRT, ZTF, PS1}


\software{
\texttt{HEASoft v6.27.2}, \texttt{XRTDAS}, \texttt{NuSTARDAS v1.9.2}, \texttt{NICERDAS v7a}, \texttt{XSPEC v12.11.0} \citep{Arnaud1996}, \texttt{IRAF} \citep{Tody1986}, \texttt{DAOPHOT} \citep{Stetson1987}, \texttt{SExtractor} \citep{Bertin1996}, \texttt{Swarp} \citep{Bertin2010}, \texttt{SCAMP} \citep{Bertin2006}, \texttt{astrometry.net} \citep{Lang2010}, Stingray \citep{Huppenkothen2019}, Astropy \citep{astropy}}





\bibliography{magnetar-bib}{}
\bibliographystyle{aasjournal}



\appendix
\setcounter{table}{0}
\renewcommand\thetable{\Alph{section}\arabic{table}}

\section{Log of X-ray observations}

\begin{table}[h]
\caption{Log of X-ray observations of J18219, including the orbital phase (\S \ref{sec: orbit_timing}) at the time of each observation.} 
\label{tab:Xobservations}
\centering
\begin{tabular}{lccccccc}
\hline\hline
\textbf{Start Time (UT)} & \textbf{Telescope} & \textbf{Instrument} &  \textbf{Exposure (s)} & \textbf{Orb. Phase} & \textbf{ObsID} & \textbf{Ref.} \\
\hline
2010-03-05 19:01:00 & \textit{Swift} & XRT/PC & 1332 & 0.37 &00031649001 & 1\\
2011-02-20 12:14:12 & \textit{Chandra} & HRC-I & 1190 & 0.20 &12499 & 2 \\
2012-02-15 06:50:00 & \textit{Swift} & XRT/PC & 1361 & 0.19 & 00032285001 & 1 \\
2012-02-21 16:24:00 & \textit{Swift} & XRT/WT & 3128 & 0.28 &00032285002 & 1\\
2012-02-24 16:50:00& \textit{Swift} & XRT/WT & 3744 & 0.32 &00032285003 & 1\\
2012-02-27 04:06:00 & \textit{Swift} & XRT/WT & 2952 & 0.35 &00032285004 & 1 \\
2012-03-01 15:22:00 & \textit{Swift} & XRT/WT & 1485 & 0.40 &00032285005 & 1\\
2012-03-04 13:58:00 & \textit{Swift} & XRT/WT & 3089 & 0.44 &00032285006 & 1 \\
2012-03-07 01:22:00 & \textit{Swift} & XRT/WT & 3214 & 0.47 &00032285007 & 1\\
2012-03-09 23:59:00 & \textit{Swift} & XRT/WT & 2995 & 0.52 &00032285008 & 1 \\
2012-03-13 01:45:00 & \textit{Swift} & XRT/WT &3249 & 0.56 &00032285009 & 1\\
2012-03-17 18:17:00 & \textit{Swift} & XRT/WT & 3349 & 0.62 &00032285010 & 1\\
2012-03-19 18:24:00 & \textit{Swift} & XRT/WT & 2819 & 0.65 &00032285011 & 1 \\
2012-10-22 09:42:59 & \textit{Swift} & XRT/PC & 558 & 0.62 &00044173001 & This work \\
2012-10-22 17:40:59 & \textit{Swift} & XRT/PC & 461 & 0.62 &00044172001 & This work \\
2017-07-22 17:27:57 & \textit{Swift} & XRT/PC & 4642 & 0.56 &00087421001 & This work\\
2018-02-28 11:15:57 & \textit{Swift} & XRT/PC & 381 & 0.61 &00087421003 & This work\\
2020-03-05 04:03:35 & \textit{Swift} & XRT/PC & 4512 & 0.76 & 03110746001 & This work \\
2020-03-06 21:32:36 & \textit{Swift} & XRT/PC & 1764 & 0.79 &03110747001 & This work\\
2020-03-11 12:58:36 & \textit{Swift} & XRT/PC & 3109 & 0.85 &03110747002 & This work \\
2020-03-15 16:31:09 & \textit{NuSTAR} & FPMA/B & 23000 & 0.91 &90601309002 & This work\\
2020-05-30 21:21:36 & \textit{Swift} & XRT/PC & 456 & 0.96 &03110746003 & This work\\
2020-06-03 07:15:34 & \textit{NICER}  & XTI & 2344 & 0.01 &3201610101 & This work\\
2020-10-20 00:40:35 & \textit{Swift} & XRT/PC & 4643 & 0.92  &03110855001 & This work\\
2021-03-09 02:22:35 & \textit{Swift} & XRT/PC & 391 & 0.86 &03110855002 & This work \\
2021-05-02 11:18:40 & \textit{NICER}  & XTI & 1189 & 0.61 &4201610101 & This work \\
\hline\hline
\end{tabular}
  \tablecomments{References: [1] \citet{LaParola2013}, [2] \citet{Karasev2012}}
\end{table}

\end{document}